\documentclass[]{spie}  %>>> use for US letter paper
\usepackage{amsmath,amsfonts,amssymb}
\usepackage{graphicx}
\usepackage[colorlinks=true, allcolors=blue]{hyperref}

\usepackage{subcaption}
\title{GALI Localization of Off-Grid GRB Directions\\ \Large{\textbf{A Comparison of Rectangular and Triangular Calibration Meshes}}}

\author[a]{Julia Saleh-Natur}
\author[a]{Ehud Behar} 
\author[a]{Omer Reich}
\author[a]{Shlomit Tarem}
\author[a]{Zvika Tarem}
\author[a]{Alex Vdovin}
\author[a]{Amir Feigenboim}
\author[a]{Roey Evental}
\author[a]{Solomon Margolin}

\affil[a]{Department of Physics, Technion - Israel Institute of Technology, Haifa, Israel}
\authorinfo{Further author information: J.S.N. E-mail: Julia.s@campus.technion.ac.il }
\begin{document} 
\maketitle

\begin{abstract}
Rapid localization of gamma-ray bursts (GRBs) is essential for multi-wavelength and multi-messenger follow-up observations, yet achieving both wide sky coverage and accurate localization remains a significant challenge. GALI (GAmma-ray burst Localizing Instrument) is a $\gamma$-ray detector concept that determines source directions using directional count-rate modulations produced by internal shadowing within a three-dimensional scintillator array. In this work, we investigate a triangular calibration mesh based on an icosphere tessellation of the hemisphere and compare its localization performance with that of the rectangular calibration mesh used in previous studies. The triangular mesh provides a more isotropic sampling of the hemisphere and improves interpolation between calibration points. Laboratory measurements performed with the full-scale GALI detector demonstrate an average angular separation of $0.86^\circ \pm 0.14^\circ$ for random source directions using the triangular calibration mesh, compared with $1.19^\circ \pm 0.28^\circ$ for the rectangular mesh. Monte Carlo simulations of 50,000 randomly distributed GRBs under realistic Low Earth Orbit background conditions, each with 1000 detected counts, yield average angular separations of $1.91^\circ \pm 1.20^\circ$ and $2.73^\circ \pm 2.18^\circ$ for the triangular and rectangular calibration meshes. The superior localization performance measured in the laboratory is attributed, in part, to higher counting statistics, with the measurements containing approximately 3--4 times as many detected counts as the simulated bursts. Additional simulations show sensitivity from 10\,keV to $\sim$1\,MeV, with a peak effective area of approximately 130 \,cm$^2$ near 100--150\,keV. Simulations of a compact 1U GALI implementation further demonstrate degree-scale localization within CubeSat volume constraints, highlighting the scalability of the concept for future GRB and multi-messenger missions.

\end{abstract}


\section{INTRODUCTION}
\label{sec:intro}  
\subsection{Why GRB Localization Still Matters}

\paragraph{} GRBs are among the most energetic transient phenomena in the Universe and continue to play a central role in high-energy and multi-messenger astrophysics \cite{Meszaros2006, Kumar2015}. Rapid and accurate localization of GRBs is essential for enabling prompt follow-up observations across X-ray, optical, infrared, and radio wavelengths, allowing the identification of afterglows, host galaxies, and the physical origin of the burst \cite{Barthelmy2005, Fruchter2006}.

Precise localization is particularly important in the era of multi-messenger astronomy. Joint detections of GRBs and gravitational-wave events hold the key to understanding neutron-star mergers. The association of GRB 170817A with the binary neutron star merger GW170817 demonstrated the scientific importance of combining electromagnetic and gravitational-wave observations \cite{Abbott2017a}. Efficient follow-up observations require localization regions that are sufficiently small for optical and X-ray facilities to rapidly identify transient counterparts.

Current GRB localization techniques generally involve a trade-off between localization accuracy and sky coverage. Coded-mask instruments, such as Swift/BAT \cite{Barthelmy2005}, achieve excellent angular precision but typically operate within a limited field of view, while distributed-detector systems such as Fermi/GBM \cite{Meegan2009} provide nearly all-sky coverage at the expense of localization accuracy. Developing compact instruments that combine wide sky coverage with improved localization performance remains an important objective for future GRB missions.

This work focuses on localization reconstruction studies using the GALI concept, with particular emphasis on random off-grid source localization and on comparing two independent sky-mesh calibration approaches based on rectangular and triangular grids. The present study expands upon the results reported in \cite{saleh2025} by introducing a triangular calibration mesh, investigating localization performance for fully random off-grid source positions, and evaluating the reconstruction method through laboratory measurements and Monte Carlo simulations.

\subsection{Current GRB Localization Methods}
\label{subsec:methods}

\begin{figure} [h!]
    \centering
    \includegraphics[width=0.75\linewidth]{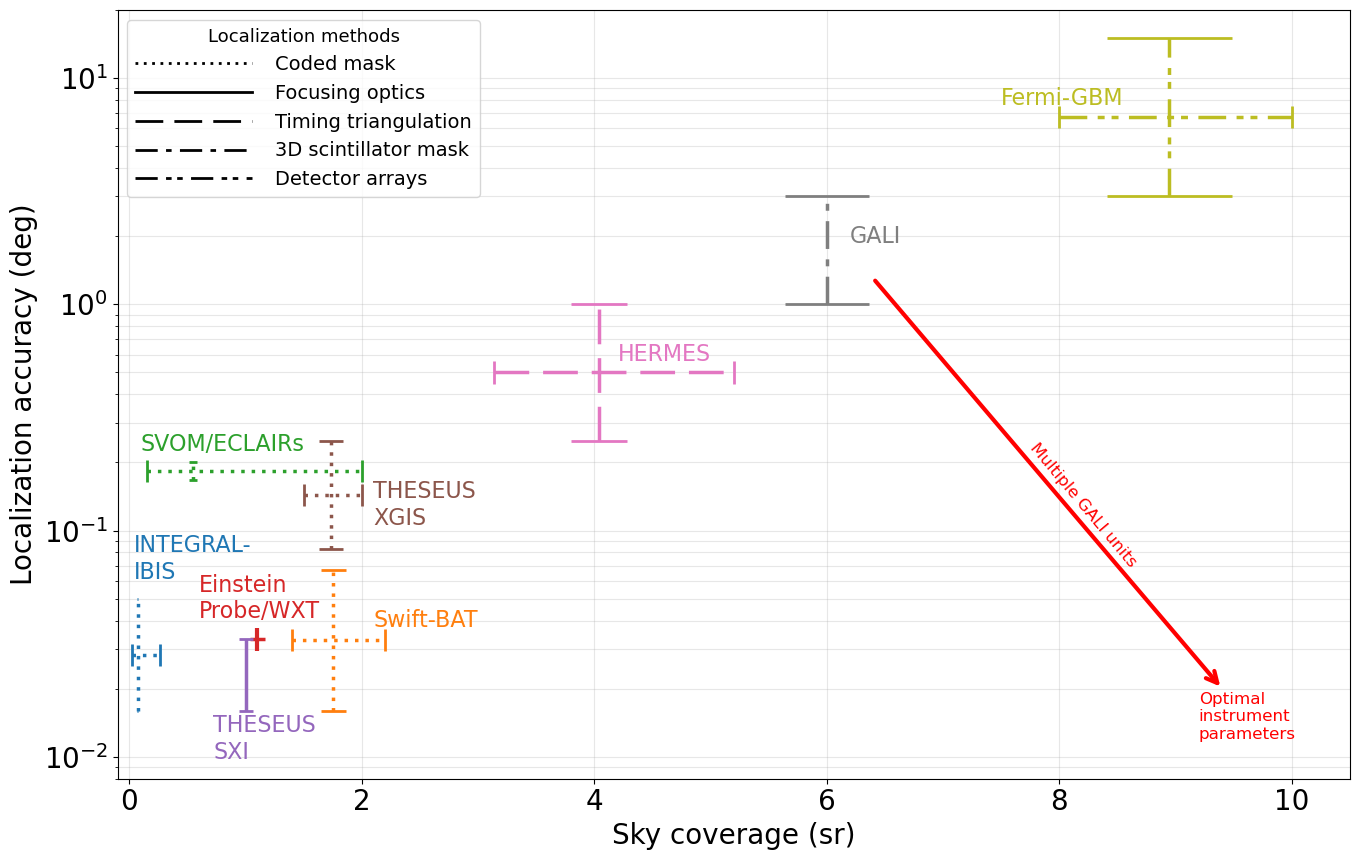}
    \caption{Comparison of current and proposed GRB localization approaches in terms of localization accuracy and sky coverage. Line styles indicate different localization methods. Coded-mask and focusing-optics instruments provide arcminute-scale localization over relatively limited fields of view. In contrast, detector-array and triangulation-based systems achieve broader sky coverage at the expense of reduced localization precision. GALI occupies an intermediate region, combining approximately hemispherical $\sim 2\pi$\,sr sky coverage with degree-scale localization capability. The red arrow illustrates a possible evolution toward multiple GALI-unit configurations and optimized instrument parameters. Plotted ranges represent reported performance intervals from the literature.}
    \label{fig:methods}
\end{figure}

\paragraph{} Figure \ref{fig:methods} compares the localization accuracy and sky coverage of several current and proposed GRB localization approaches. Coded-mask instruments (dotted lines), such as Swift/BAT \cite{Barthelmy2005}, INTEGRAL/IBIS \cite{ubertini2003}, SVOM/ECLAIRs \cite{Wei2016}, and THESEUS/XGIS \cite{Amati2018}, provide arcminute-scale localization accuracy, but generally over a limited field of view. Focusing-optics instruments (solid lines), such as Einstein Probe/WXT \cite{Yuan2015} and THESEUS/SXI \cite{Amati2018}, can also provide arcminute-level localization over wide soft-X-ray fields of view. In contrast, Fermi/GBM \cite{Meegan2009} uses relative count rates between scintillator detector units (dash-dot-dot lines) to provide very large sky coverage, but with localization accuracy typically of several degrees. Triangulation-based systems (dashed lines), such as the HERMES initiative \cite{sanna2020}, obtain localization from arrival-time differences measured by multiple spacecraft. This approach can provide wide sky coverage, but it depends on a distributed detector constellation.

GALI (dash-dot lines) occupies an intermediate region in this parameter space. It provides approximately hemispherical sky coverage and degree-scale localization using a compact, single-platform detector. Figure \ref{fig:methods} therefore illustrates how GALI bridges the gap between high-precision but narrower-field instruments and very wide-field systems. The modular nature of the GALI concept may also allow future implementations with larger detector volumes or multiple units, improving sensitivity and field of view while preserving the same localization principle. The red arrow in Figure \ref{fig:methods} schematically indicates a possible development path toward configurations employing multiple GALI units and optimized instrument parameters, potentially enabling increased sky coverage and improved localization performance.

\section{GALI}
\label{sec:GALI}

\subsection{Concept}
\label{subsec:concept}

\paragraph{} GALI (GAmma-ray burst Localizing Instrument) is a compact GRB localization concept based on a three-dimensional scintillator architecture. Its localization capability originates from angle-dependent internal masking between numerous small scintillators distributed within a confined detector volume. As incoming $\gamma$-rays propagate through the detector, different source directions produce distinct relative count distributions across the scintillator array. These directional response patterns enable reconstruction of the incoming photon direction through statistical comparison with calibrated detector tables. For a comprehensive description of the instrument, see Saleh-Natur (2025) et al. \cite{saleh2025}. 

\begin{figure} [h!]
    \centering
    \includegraphics[width=0.7\linewidth]{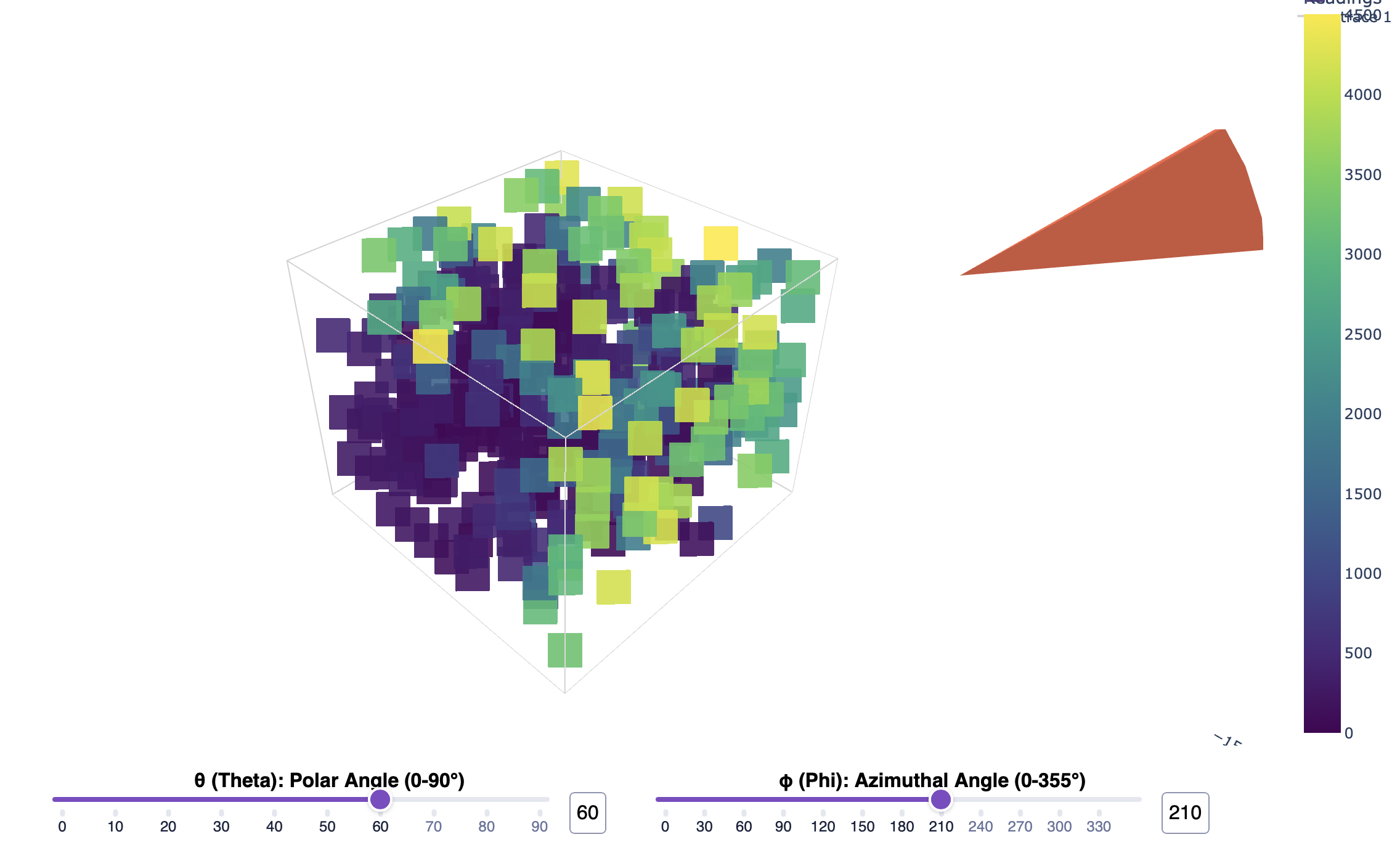}
    \caption{Measured detector response of the GALI scintillator array for a source located at $\theta = 60^\circ$ and $\phi = 210^\circ$. The colored cubes represent the relative detector readings normalized to counts per second, as indicated by the color bar. The response map is obtained from laboratory measurements using a 120\,s exposure. The orange arrow indicates the direction of the incoming $\gamma$-ray source. Direction-dependent count distributions arise from internal shadowing within the three-dimensional scintillator array and form the basis of the GALI localization method.}
    \label{fig:concept}
\end{figure}

Unlike conventional coded-mask instruments, GALI does not require an external mask to encode directional Information. Instead, the scintillators themselves act simultaneously as both detecting and masking elements. This allows the instrument to maintain a wide field of view while preserving directional sensitivity within a compact detector geometry.
Figure \ref{fig:concept} presents an example of the measured detector response for a source located at $\theta = 60^\circ$ and $\phi = 210^\circ$. The colored cubes represent the relative detector readings in counts per second, obtained from laboratory measurements using 120\,s exposures to a $\gamma$-source. The orange arrow indicates the direction of the incoming photons. 
Internal shadowing within the scintillator array produces a unique distribution of detector count rates for each source direction. As illustrated, scintillators on the side opposite the source record lower count rates than those facing the source. GALI reconstructs the source direction by statistically comparing the measured detector response with calibrated detector response maps, enabling degree-scale localization and the determination of localization uncertainty contours on the sky.

\subsection{Operating Model}
\label{subsec:operating_model}

\begin{figure} [h!]   
    \centering
    \includegraphics[width=0.85\linewidth]{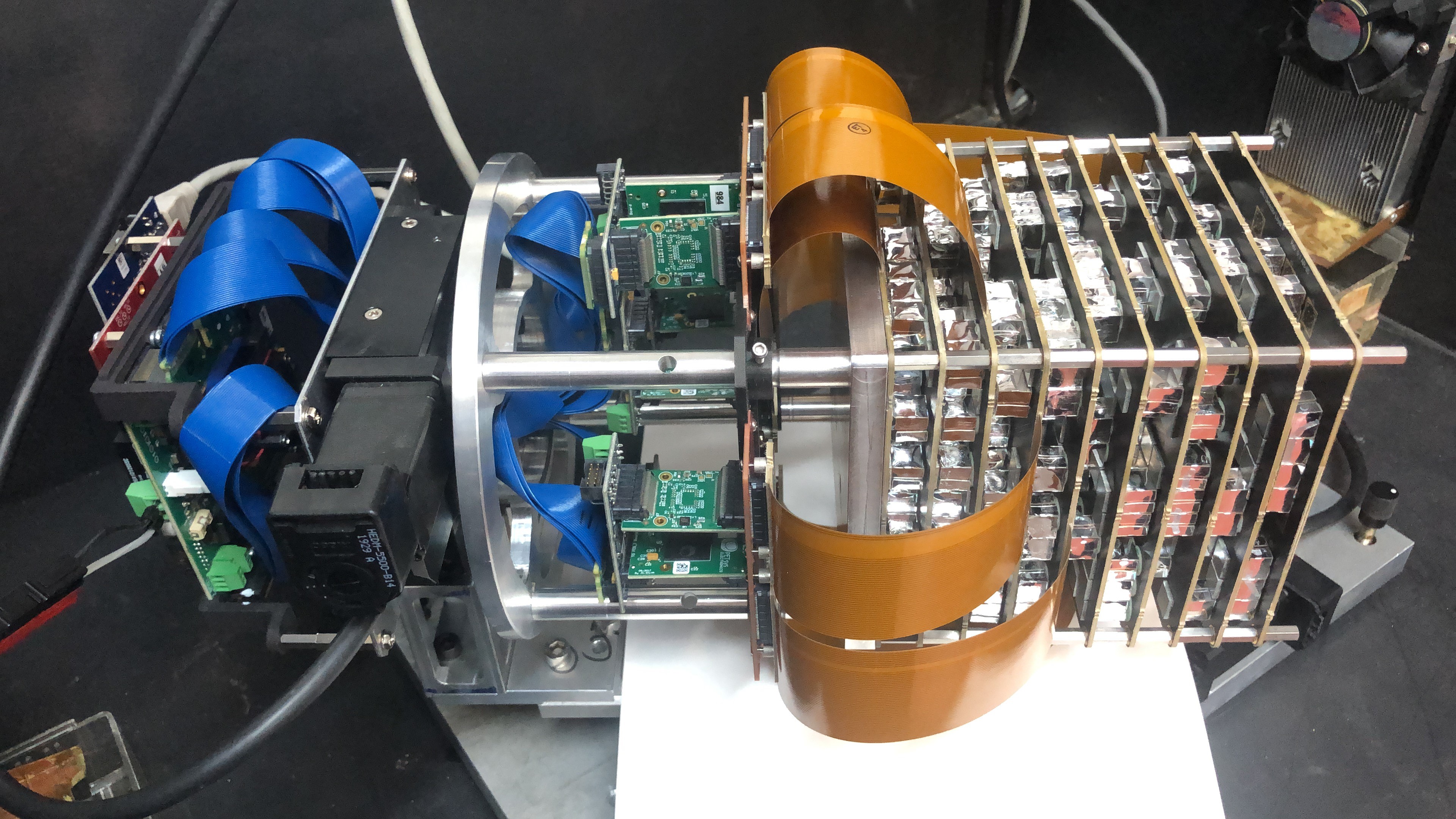}
    \caption{Current GALI detector model consisting of 362 CsI(Tl) scintillators randomly distributed across nine detector layers. Each scintillator is mounted on an individual SiPM and read out by a single ASIC channel. The detector is installed on two rotating stages for the localization tests and set in a thermally insulated box.}
    \label{fig:GALI}
\end{figure}

\paragraph{} Figure \ref{fig:GALI} presents the current GALI detector model and its internal architecture. The detector assembly consists of 362 cubic CsI(Tl) scintillators, each with dimensions of $(9 \times 9 \times 9)$\,mm$^3$, randomly distributed across nine polyimide printed circuit boards (PCBs). The random three-dimensional arrangement of the scintillators is designed to maximize directional diversity in the detector response while preserving a compact instrument geometry. Each scintillator is coated with a SiO$_2$ protective layer to reduce sensitivity to humidity and wrapped with reflective polymer foil to improve light-collection efficiency and suppress optical crosstalk between neighboring detectors. The scintillators are optically coupled to Hamamatsu S14160-6050HS silicon photomultipliers (SiPMs), which convert the scintillation light generated by $\gamma$-ray interactions into electrical signals.

To minimize obstruction of incoming $\gamma$-rays, the detector electronics are positioned beneath the scintillator volume and connected via flexible cables. The readout system is based on TOFPET application-specific integrated circuits (ASICs), where each scintillator is connected to an individual readout channel. Detector signals are processed by a multi-processor system-on-chip incorporating FPGA-based and embedded computing capabilities, enabling simultaneous acquisition of pulse-height and timing information from all detector channels.

To mitigate radiation-induced degradation of the SiPM performance in space environments, copper shielding is incorporated around and beneath the SiPM assemblies. The shielding is designed to reduce proton-induced increases in dark current during long-duration operation in orbit. The electronics module is additionally enclosed within an aluminum radiation shield.

\section {Localization Reconstruction Algorithms and Calibration Meshes}
\label{sec:cstat}

\paragraph{} The GALI localization method reconstructs the source direction by comparing the measured photon counts across the scintillator array with the expected detector response for different source directions. Because photon detection is inherently a counting process, the detected counts in each scintillator are described by Poisson statistics. This statistical description is particularly important for short GRBs, where the brief burst duration results in relatively few detected photons.

For a given source direction, the expected detector response is described by the response table $m_i$, while the measured counts in each scintillator are denoted by $S_i$. The source direction is reconstructed by maximizing the likelihood between the measured and expected count distributions over the detector array. Assuming Poisson statistics, the likelihood function is given by

\begin{equation}
L=\prod_{i=1}^{N}
\frac{(t m_i)^{S_i}}{S_i!}
\exp(-t m_i),
\label{eq:likelihood}
\end{equation}

\noindent where $N$ is the number of scintillators, and $t$ is the exposure time. Taking the logarithm of Eq. \ref{eq:likelihood} and applying the Stirling approximation yields the Cash statistic (Cstat) \cite{Cash1979}:

\begin{equation}
C =
2 \sum_{i=1}^{N}
\left[
(tm_i)-S_i+
S_i\ln\left(\frac{S_i}{tm_i}\right)
\right].
\label{eq:cstat}
\end{equation}

\noindent The reconstructed source direction corresponds to the sky position that minimizes the Cstat value.

\begin{figure} [h!]
    \centering
    \includegraphics[width=0.9\linewidth]{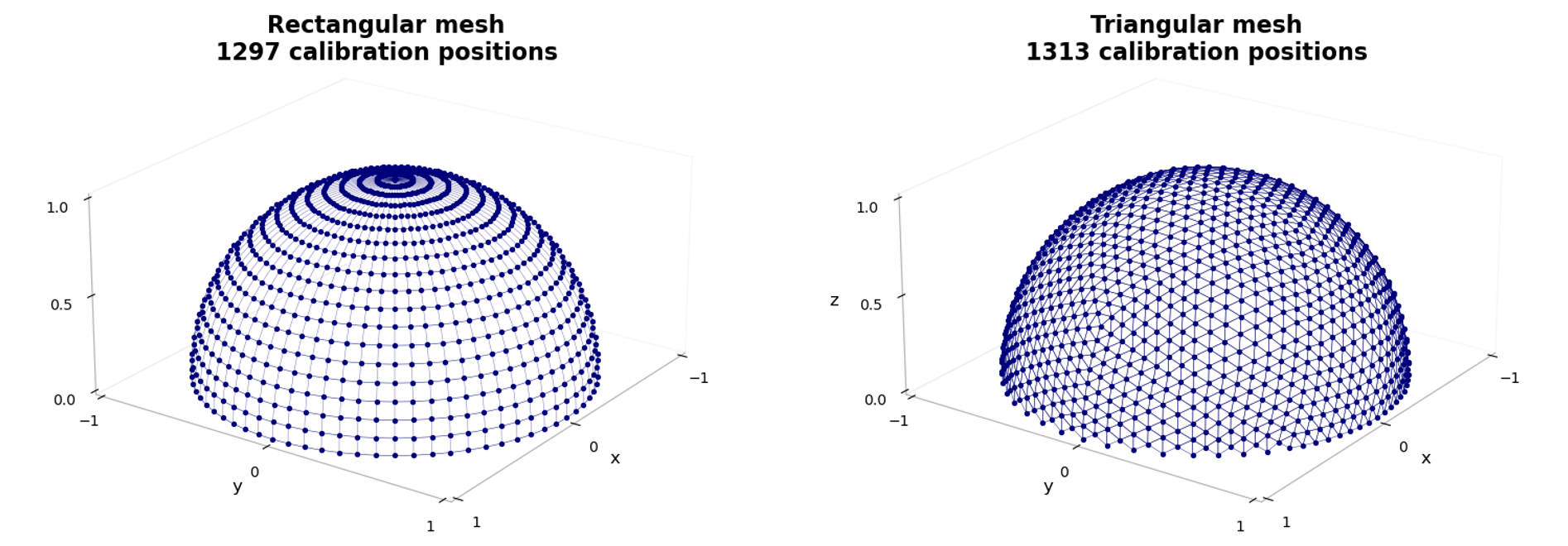}
     \caption{Schematic representation of the rectangular and triangular calibration meshes used in this work. The rectangular mesh (left) employs a regular $5^\circ \times 5^\circ$ grid over the hemisphere, while the triangular mesh (right) is generated using an icosphere-based tessellation, producing an approximately uniform triangular sampling with a comparable number of calibration positions. Blue markers denote the calibration positions.}    
     \label{fig:mesh}
\end{figure}

The expected detector response, $m_i$, is obtained from calibration measurements on a discrete angular mesh. The sampling geometry of this mesh determines the detector response table and its interpolation during localization reconstruction. In this work, two calibration mesh configurations for sampling the hemisphere are investigated: a rectangular grid and a triangular grid, as shown in Figure \ref{fig:mesh}. The two meshes are compared to evaluate the effect of the calibration sampling strategy on localization performance. For the rectangular mesh, calibration measurements are performed on $5^\circ \times 5^\circ$ a grid across the hemisphere, resulting in 1297 calibration positions. For the triangular mesh, the calibration positions are generated using an icosphere-based tessellation of the hemisphere \cite{Lee2017}, producing 1313 calibration positions distributed approximately uniformly over the field of view. Compared with the rectangular grid, the triangular tessellation provides more isotropic angular sampling and a more uniform distribution of nearest-neighbor distances across the hemisphere. In both cases, long-exposure measurements of 120\,s are used to construct the detector response tables.

To evaluate the localization performance of each calibration mesh, measurements are acquired at random off-grid source directions that do not coincide with the corresponding calibration points. At each direction, 100 short exposures of 0.5\,s are measured.

For each short exposure, the Cstat value is first evaluated over the full calibration mesh, and the mesh point with the minimum Cstat value is used as an initial localization estimate. To improve the angular precision beyond the discrete mesh spacing, a local interpolation of the Cstat surface is then performed around the best-fit mesh position. The interpolated Cstat distribution is subsequently minimized on a finer sub-grid to obtain the final reconstructed direction. Dedicated interpolation schemes are implemented separately for the rectangular and triangular mesh configurations.

The angular localization accuracy is quantified using the angular separation $\alpha$ between the reconstructed and source directions:

\begin{equation}
\alpha =
\arccos
\left[
\cos\theta_1 \cos\theta_2 +
\sin\theta_1 \sin\theta_2
\cos(\phi_1-\phi_2)
\right],
\label{eq:alpha}
\end{equation}

\noindent where $(\theta_1,\phi_1)$ denotes the true source direction and $(\theta_2,\phi_2)$ denotes the reconstructed direction. The quantity $\alpha_{\mathrm{avg}}$ represents the average angular deviation obtained from the 100 short-exposure measurements at a given source position. In contrast, $\langle \alpha_{\mathrm{avg}} \rangle$ denotes the subsequent average over all measured sky positions.

\section{Laboratory Localization Results for off-grid Source Directions}
\label{sec:lab results}

\subsection{Experimental Setup and Test}
\label{subsec:setup_test}

\paragraph{} To evaluate the localization capability and angular resolution of GALI, the detector is mounted on a two-axis rotation platform driven by two independent motors (see Figure \ref{fig:GALI}). The system enabled controlled scanning over the hemisphere, covering angular ranges of $0^\circ \leq \theta \leq 90^\circ$ and $0^\circ \leq \phi \leq 360^\circ$. The detector assembly and readout electronics are enclosed in an optically and thermally controlled box to suppress stray light and stabilize operating conditions during measurements.

The detector is exposed to a 10\,mCi $^{241}$Am source positioned approximately 4.57\,m away from the detector center, providing a nearly point-like distant $\gamma$-ray source for localization measurements. The analysis focused on the 59.6\,keV emission line of $^{241}$Am. The short 0.5\,s bursts result in approximately 3500 photons on the detector.

For each scintillator, the five spectra with the highest signal across all measured directions are identified and fitted with a combined Gaussian model around the 59.6\,keV line. The resulting line center and width $\sigma$ are then used to extract source counts within a $3\sigma$ interval around the photopeak, including for viewing angles where the signal is weak. These extracted counts are used to construct the detector response and perform the localization reconstruction analysis.

In the laboratory measurements, the background contribution is negligible compared to the source signal. For space-based GRB observations, however, background counts introduce additional statistical fluctuations that affect the localization uncertainty. In such cases, the background level will be measured on each scintillator prior to the GRB trigger and subtracted from the measured detector counts, while the corresponding Poisson uncertainty will remain. One advantage of GALI is that inner scintillators collect only a few background counts.

\subsection{Laboratory Localization Results}
\label{subsec:lab localization results}

\paragraph{} The localization performance of GALI is evaluated using measurements acquired with the triangular and rectangular calibration meshes. We first characterize the detector response using the measured effective area, and then evaluate localization performance for random off-grid source directions.

\begin{figure}[h!]
    \centering
    \includegraphics[width=0.8\linewidth]{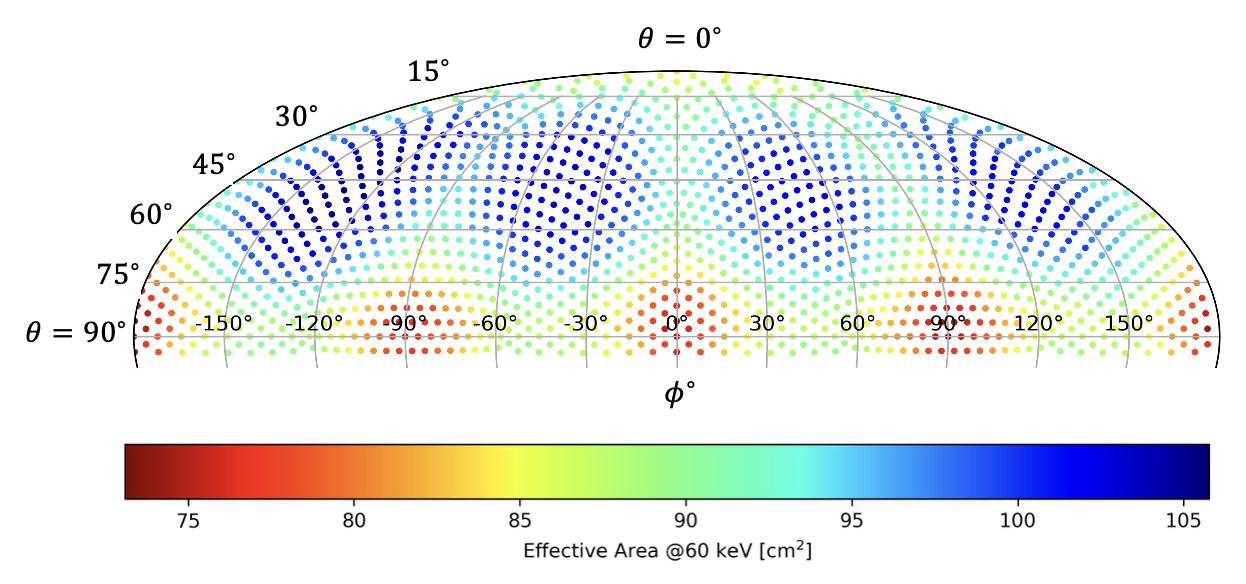}
    \caption{Measured effective area of the current GALI detector model at 59.6\,keV for the triangular calibration mesh. The map is shown in spherical coordinates $(\theta,\phi)$, where the color scale represents the effective area. Variations in the effective area arise from the directional dependence of the projected detector geometry.}
    \label{fig:skymap_tri}
\end{figure}

Figure \ref{fig:skymap_tri} presents the measured effective area of the current GALI detector model at 59.6\,keV for the triangular calibration mesh. The effective area is calculated as the ratio of the measured photon flux to the known incident photon rate. This value is corrected by 10\% to account for attenuation in the 4.57\,m air path between the source and the detector. The map is shown in spherical coordinates $(\theta,\phi)$, where each point corresponds to a calibration direction and the color scale represents the effective area in units of cm$^2$. Variations in the effective area arise from the detector's geometry and the directional dependence of its projected cross section. The largest effective area is observed near the detector corners, where the detector's projected cross-section facing the source is maximal. 

The overall effective area shown in Figure \ref{fig:skymap_tri} is slightly lower than that reported in Ref.\cite{saleh2025}, where we overcompensated for photon absorption by the foam inserted to damp mechanical vibrations. The present measurements are performed without this foam. Despite this reduction, the detector maintains a relatively high effective area across the measured hemisphere, demonstrating the wide field-of-view capability of the GALI concept.

\begin{figure}[h!]
    \centering
    \begin{subfigure}[b]{0.48\linewidth}
        \centering
        \includegraphics[width=\linewidth]{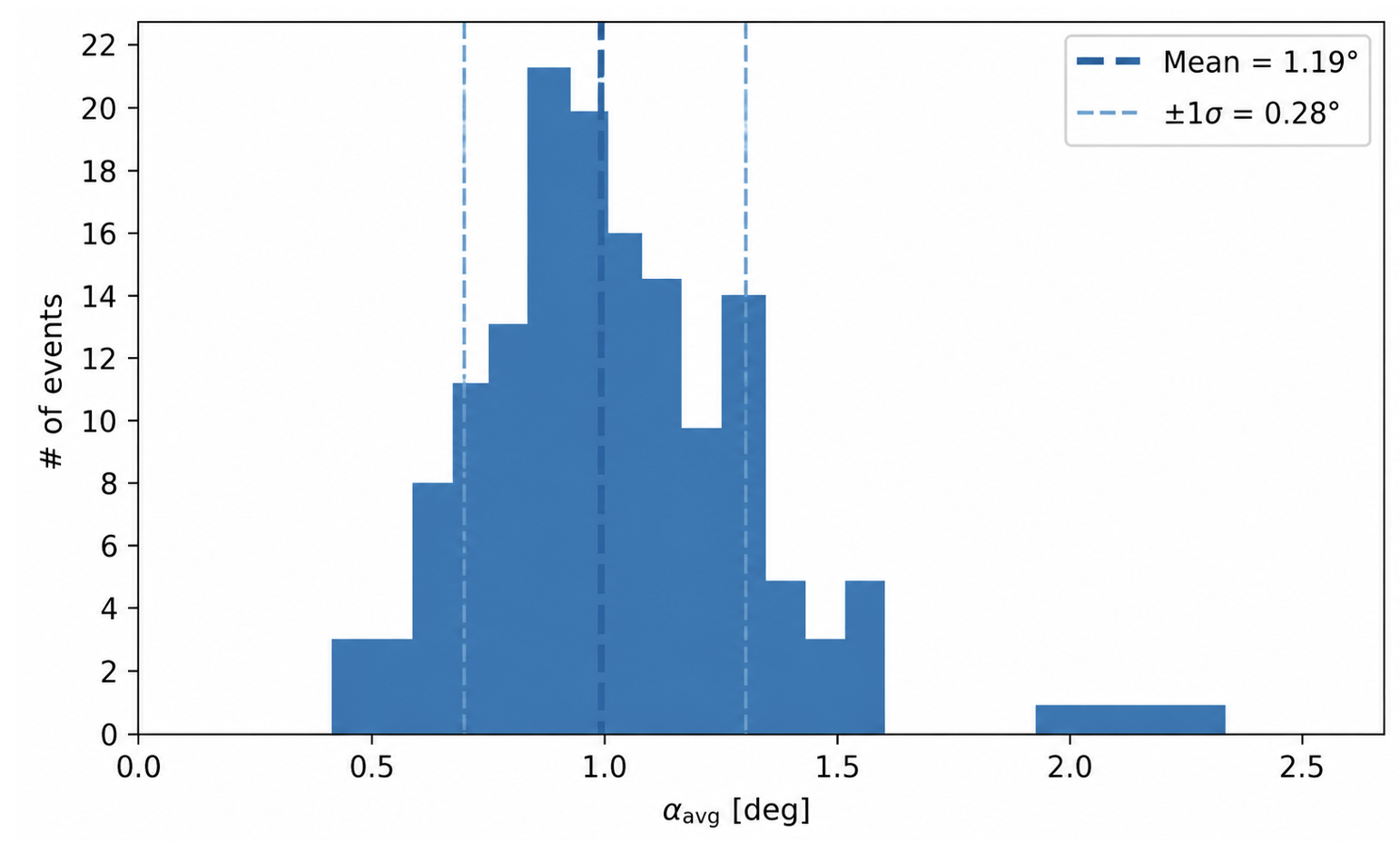}
        \caption{Rectangular mesh reconstruction}
        \label{fig:histogram_rec_random}
    \end{subfigure}
    \hfill
    \begin{subfigure}[b]{0.48\linewidth}
        \centering
        \includegraphics[width=\linewidth]{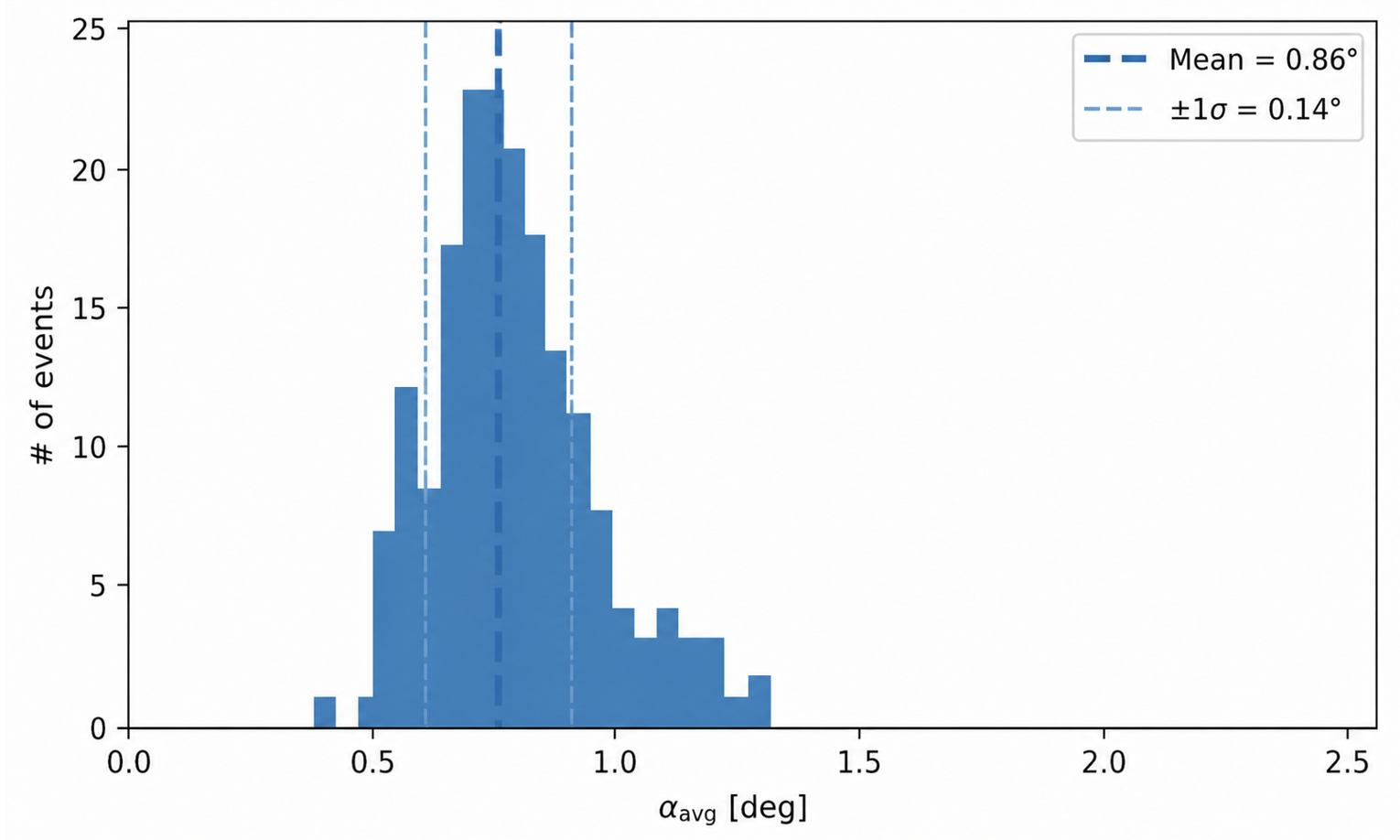}
        \caption{Triangular mesh reconstruction}
        \label{fig:histogram_tri_random}
    \end{subfigure}
    \caption{Distribution of the average angular separation, $\alpha_{\mathrm{avg}}$, using the rectangular and triangular calibration meshes. Most reconstructed events achieve sub-degree localization accuracy for both calibration geometries, demonstrating robust reconstruction performance for short 0.5\,s laboratory exposures.}
    \label{fig:histograms_random_compare}
\end{figure}

The detector response characterized above is then used to evaluate the localization performance for random off-grid source directions using both the rectangular and triangular calibration meshes. Figure \ref{fig:histograms_random_compare} presents the distributions of the average angular separation, $\alpha_{\mathrm{avg}}$, between the reconstructed and true source directions for the two calibration geometries.
Both distributions demonstrate that most reconstructed source positions achieve sub-degree localization accuracy. Using the triangular calibration mesh yields $\langle\alpha_{avg}\rangle=0.86^\circ$ with $\sigma=0.14^\circ$. The rectangular calibration mesh yields $\langle\alpha_{avg}\rangle=1.19^\circ$ with $\sigma=0.28^\circ$. Both calibration geometries achieve good localization performance, although the triangular calibration mesh consistently yields smaller angular separations than the rectangular mesh.

\begin{figure}[h!]
    \centering

    \begin{subfigure}[b]{0.48\linewidth}
        \centering
        \includegraphics[width=\linewidth]{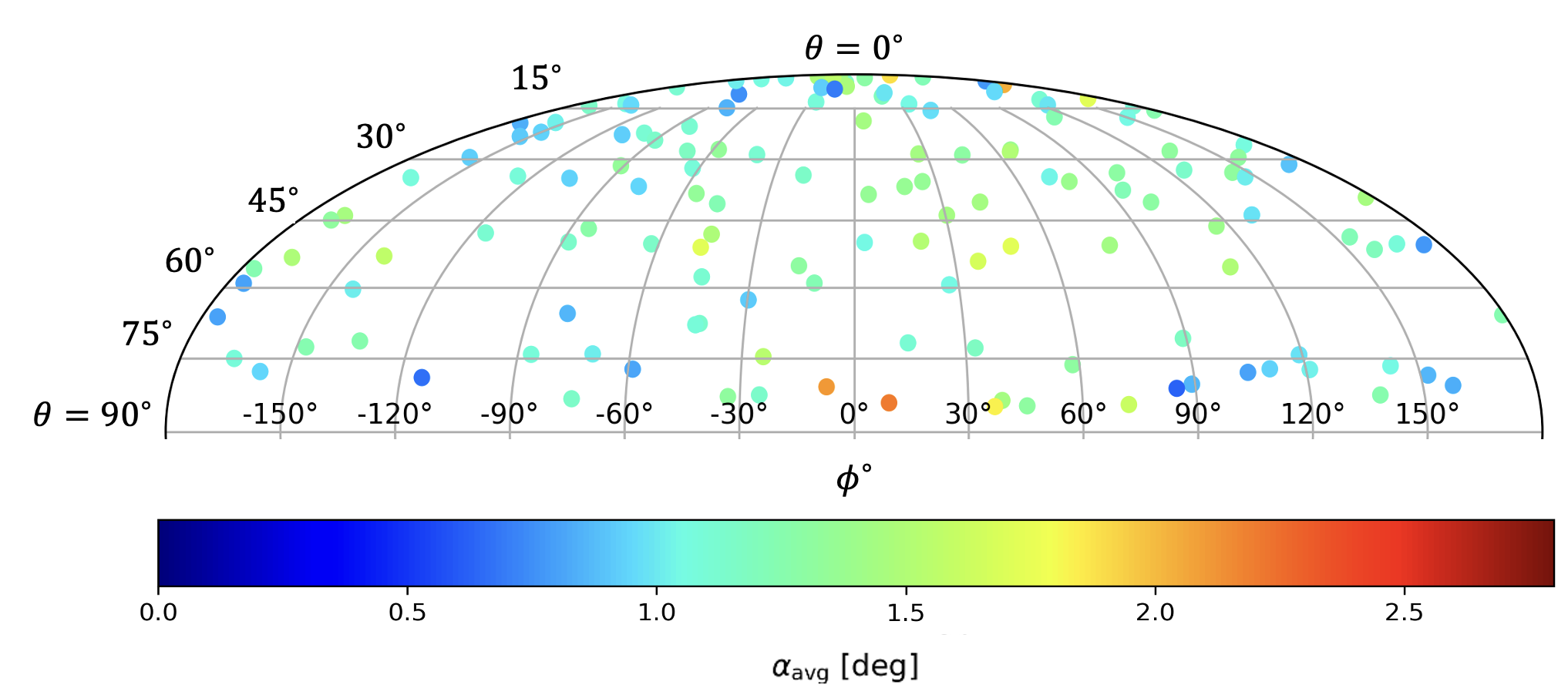}
        \caption{Rectangular mesh reconstruction}
        \label{fig:scatter_rec_random}
    \end{subfigure}
    \hfill
    \begin{subfigure}[b]{0.48\linewidth}
        \centering
        \includegraphics[width=\linewidth]{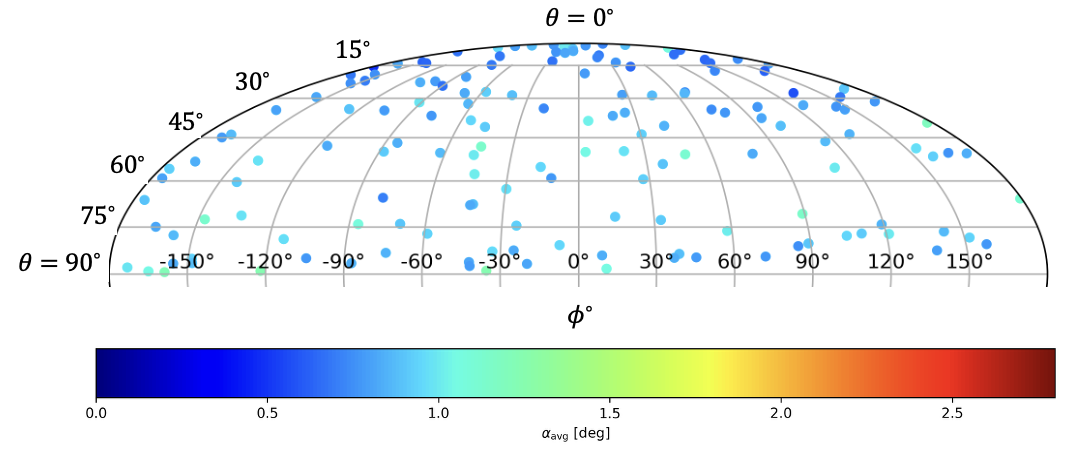}
        \caption{Triangular mesh reconstruction}
        \label{fig:scatter_tri_random}
    \end{subfigure}
    \caption{Sky distribution of the average angular separation, $\alpha_{\mathrm{avg}}$, between the reconstructed and true source off-grid directions using the rectangular (a) and triangular (b) calibration meshes. Each point corresponds to an off-grid source direction in spherical coordinates $(\theta,\phi)$, while the color scale represents the average angular separation (in degrees) over 100 reconstructions.} 
    \label{fig:scatter_random_compare}
\end{figure}

To examine how the localization performance varies across the field of view, Figure \ref{fig:scatter_random_compare} presents the sky distribution of $\alpha_{\mathrm{avg}}$ for the rectangular and triangular calibration meshes. The localization accuracy remains consistent across most of the measured hemisphere, with only slightly larger angular separations near the hemisphere boundaries. To minimize edge effects associated with the interpolation, the off-grid source directions are restricted to $\theta \leq 85^\circ$ for the rectangular calibration mesh, whose calibration measurements extend to $\theta = 90^\circ$, and to $\theta \leq 90^\circ$ for the triangular calibration mesh, whose calibration measurements extend to $\theta = 95^\circ$. This margin ensures that all reconstructed source directions remain well within the calibrated region.

\begin{figure}[h!]
    \centering
    \includegraphics[width=0.65\linewidth]{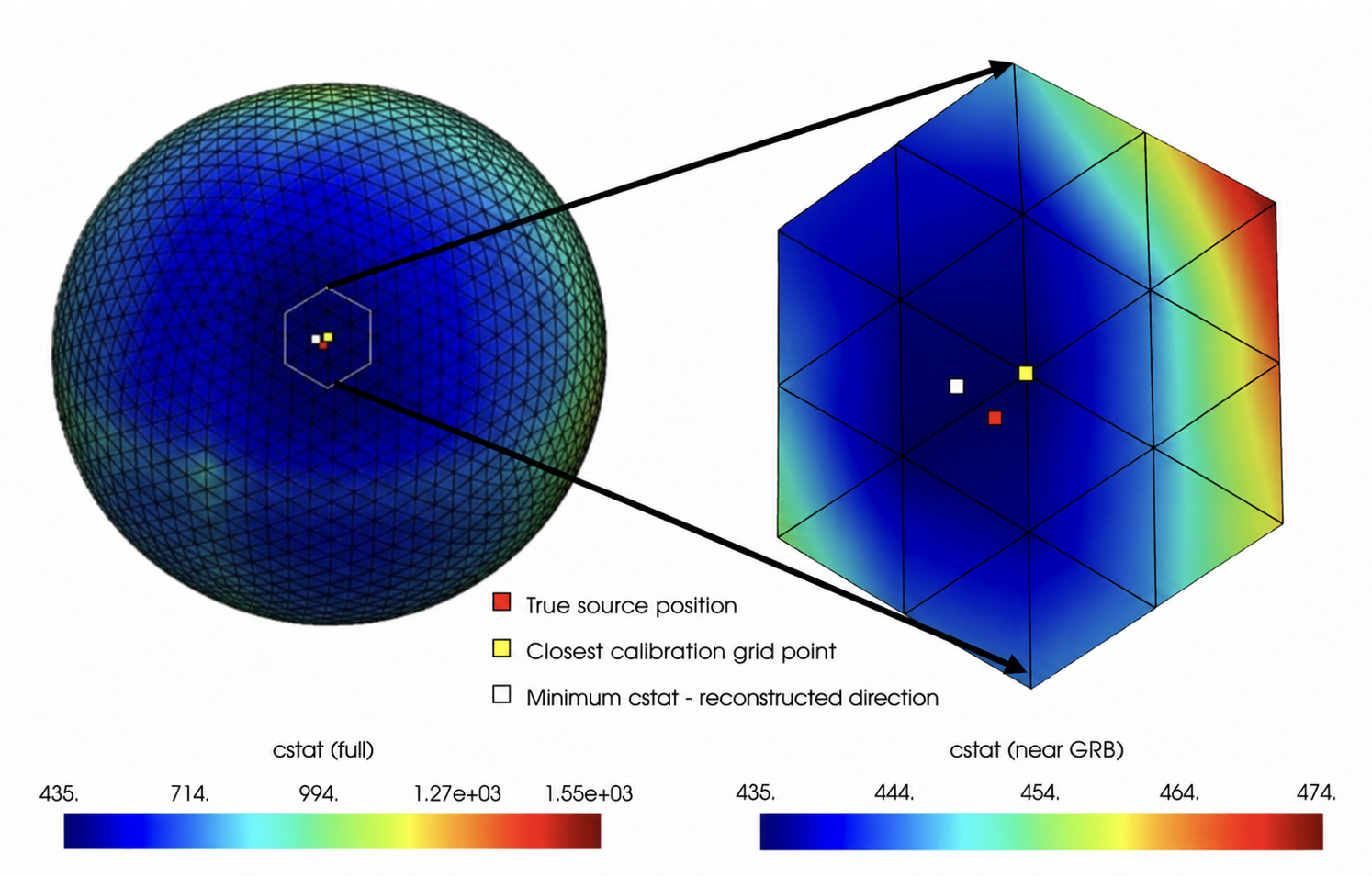}
    \caption{Localization example of an off-grid source direction using the triangular calibration mesh. The left panel shows the reconstructed Cstat map over the hemisphere, while the right panel shows a zoomed-in view around the localization minimum. The red, yellow, and white squares indicate the true source position, the closest calibration grid point, and the reconstructed source direction corresponding to the global minimum of the Cstat surface, respectively. The color scales represent the Cstat value; the zoomed-in panel uses a rescaled color range corresponding to the local variation in Cstat. For this example, $\alpha = 1.36^\circ$.}
    \label{fig:tri_reconstruction_example}
\end{figure}

Figure \ref{fig:tri_reconstruction_example} presents an example of the localization of an off-grid source direction using the triangular calibration mesh. The left panel shows the reconstructed Cstat surface over the hemisphere together with the triangular calibration mesh, while the right panel shows a zoomed-in view around the localization minimum. The red square marks the true source position, the yellow square indicates the closest calibration grid point, and the white square marks the reconstructed source direction corresponding to the global minimum of the Cstat surface. In this example, the angular separation is $\alpha = 1.36^\circ$. For clarity, the zoomed-in panel uses a separate color scale optimized for the local Cstat range around the minimum, highlighting the structure of the Cstat surface.

\section{Simulation results }
\label{sec:simulation}

\paragraph{} The laboratory measurements presented in the previous section demonstrate the localization performance of GALI using experimentally measured calibration data. However, laboratory measurements are limited to a finite set of source positions and exhibit negligible background. To further evaluate localization performance across the full hemisphere, Monte Carlo simulations are conducted using the validated GALI simulation framework developed in prior work \cite{saleh2024,saleh2025}. The simulations are carried out with the full GALI detector model implemented in the MEGAlib environment \cite{Zoglauer.soft06018Z}, which is based on the Geant4 toolkit \cite{Agostinelli2003, Allison2006, Allison2016}. The detailed detector geometry, including all detectors, shielding, and structural components affecting the instrument response, is described in Ref.~\cite{saleh2024}. 

\subsection{Localization Results for Off-Grid GRB Directions}
\label{subsec:simulation localization results}

\paragraph{} Following the same procedures used for the laboratory measurements, rectangular and triangular calibration meshes are generated over the hemisphere. High-statistics simulations are first performed at each calibration position to construct the corresponding calibration maps. Unlike the laboratory measurements, the simulations include the expected $\gamma$-ray background in a 600\,km Low Earth Orbit, as described and validated in Ref.~\cite{saleh2025}. The background contribution is simulated independently and subtracted from the detector counts prior to the localization analysis. Subsequently, a total of 50,000 GRBs with a duration of 1\,s and a photon flux of $10~\gamma~\mathrm{cm}^{-2}\,\mathrm{s}^{-1}$ are simulated from random off-grid directions uniformly distributed across the hemisphere. The burst spectra are modeled using Band-function parameters with low- and high-energy photon indices of $-1.1$ and $-2.3$, respectively, and a break energy of $E_0=266$\,keV \cite{band1993}. For each burst, the detector count distribution is compared with the corresponding calibration table using the Cstat-based reconstruction algorithm described in Section~\ref{sec:cstat}, and the source direction is reconstructed using the interpolation procedure associated with each calibration mesh.

\begin{figure}[h!]
    \centering
    \begin{subfigure}[b]{0.48\linewidth}
        \centering
        \includegraphics[width=\linewidth]{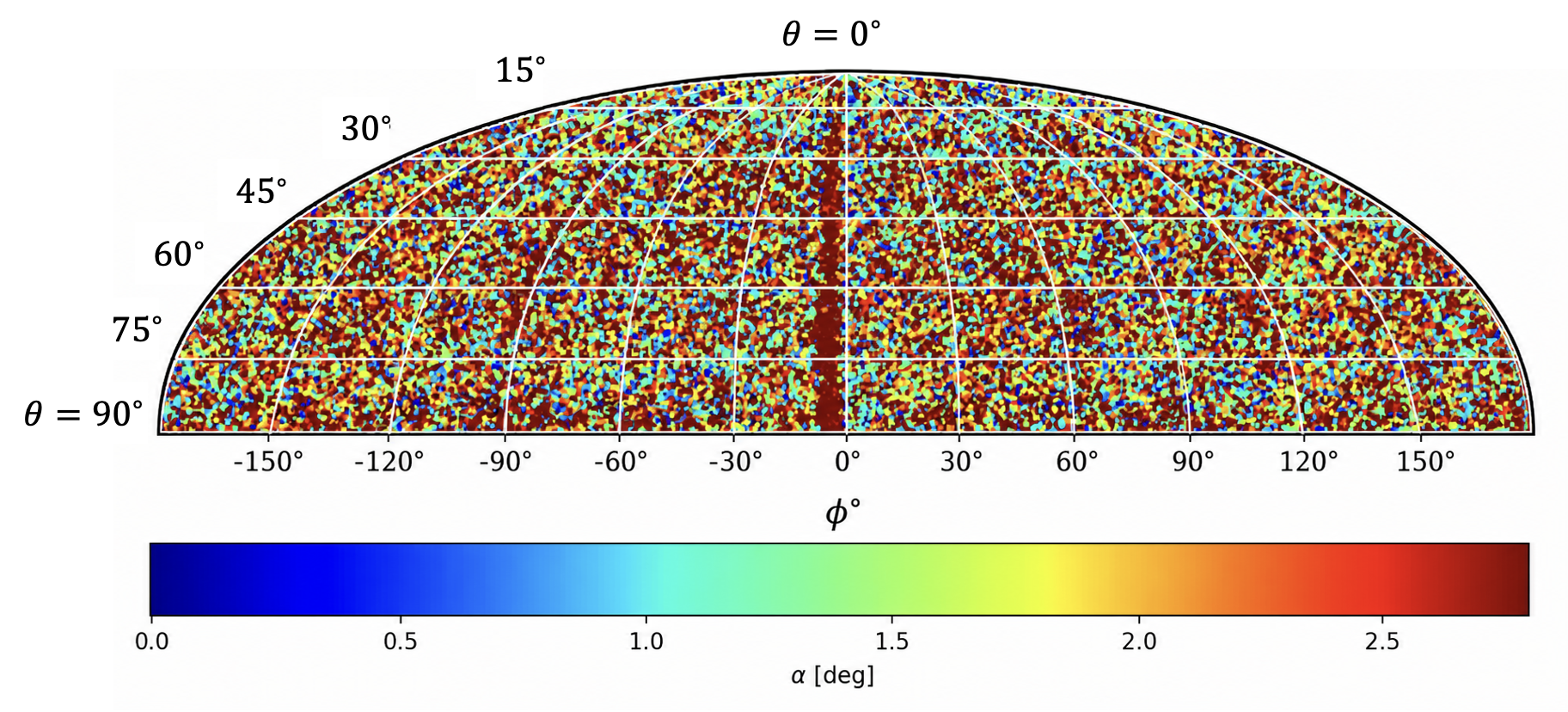}
        \caption{Rectangular mesh reconstruction}
        \label{fig:alpha_rec_simulations}
    \end{subfigure}
    \hfill
    \begin{subfigure}[b]{0.48\linewidth}
        \centering
        \includegraphics[width=\linewidth]{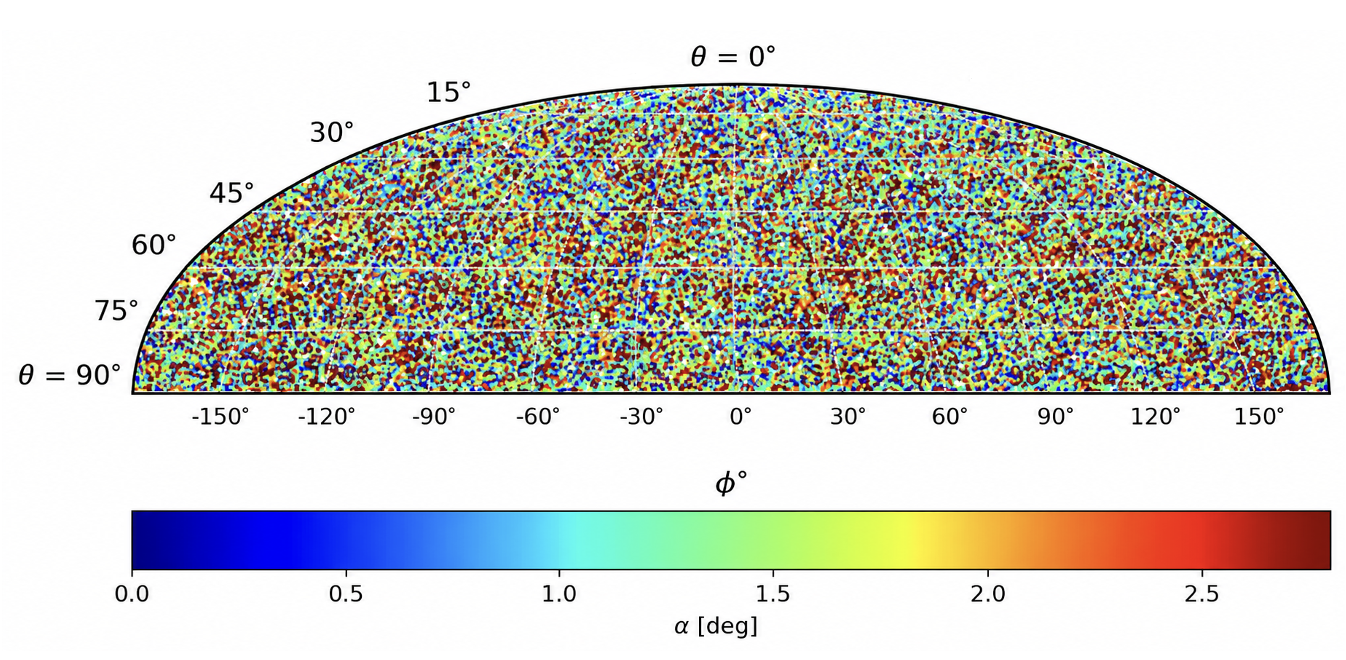}
        \caption{Triangular mesh reconstruction}
        \label{fig:alpha_tri_simulations}
    \end{subfigure}
    \caption{Sky distribution of the angular separation, $\alpha$, for 50,000 simulated GRBs reconstructed using the rectangular (a) and triangular (b) calibration meshes. The simulated GRBs follow a Band-function spectrum and are uniformly distributed over the hemisphere. The color scale is identical to that used in Figure \ref{fig:scatter_random_compare}, enabling direct comparison between the laboratory and simulation results.}
    \label{fig:alpha_simulations}
\end{figure}

Figure~\ref{fig:alpha_simulations} presents the sky distribution of the angular separation, $\alpha$, obtained for the simulated GRBs reconstructed using the rectangular and triangular calibration meshes. Both calibration geometries provide relatively uniform localization performance across the hemisphere. In the rectangular-mesh reconstruction, slightly larger angular separations appear near $\phi=0^\circ$, due to the azimuthal boundary of the rectangular calibration map, which does not appear in the triangular map. To facilitate direct comparison with the laboratory measurements, the same color scale as in Figure~\ref{fig:scatter_random_compare} is used.

\begin{figure}[h!]
    \centering
    \begin{subfigure}[b]{0.48\linewidth}
        \centering
        \includegraphics[width=\linewidth]{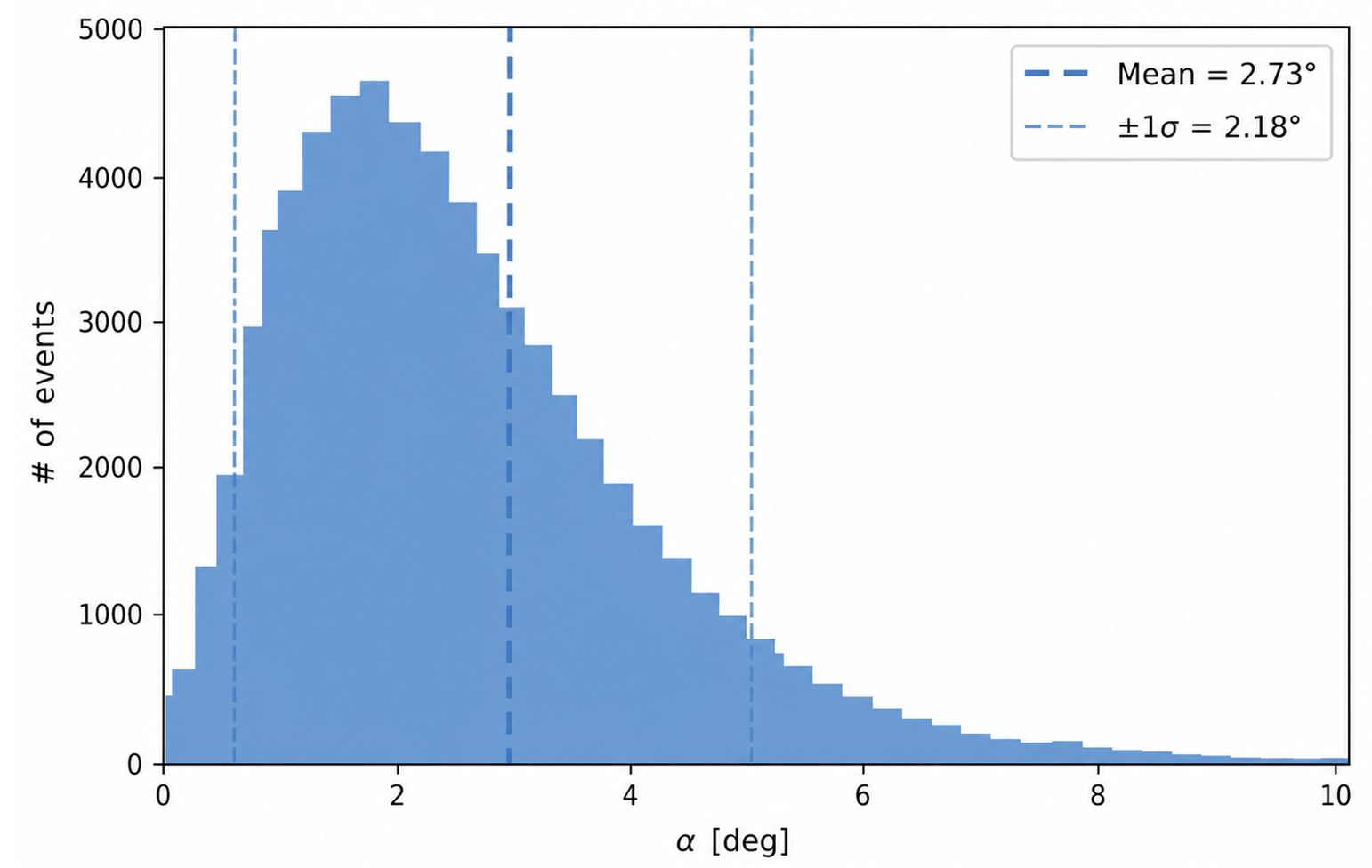}
        \caption{Rectangular mesh reconstruction}
        \label{fig:hist_rec_simulations}
    \end{subfigure}
    \hfill
    \begin{subfigure}[b]{0.48\linewidth}
        \centering
        \includegraphics[width=\linewidth]{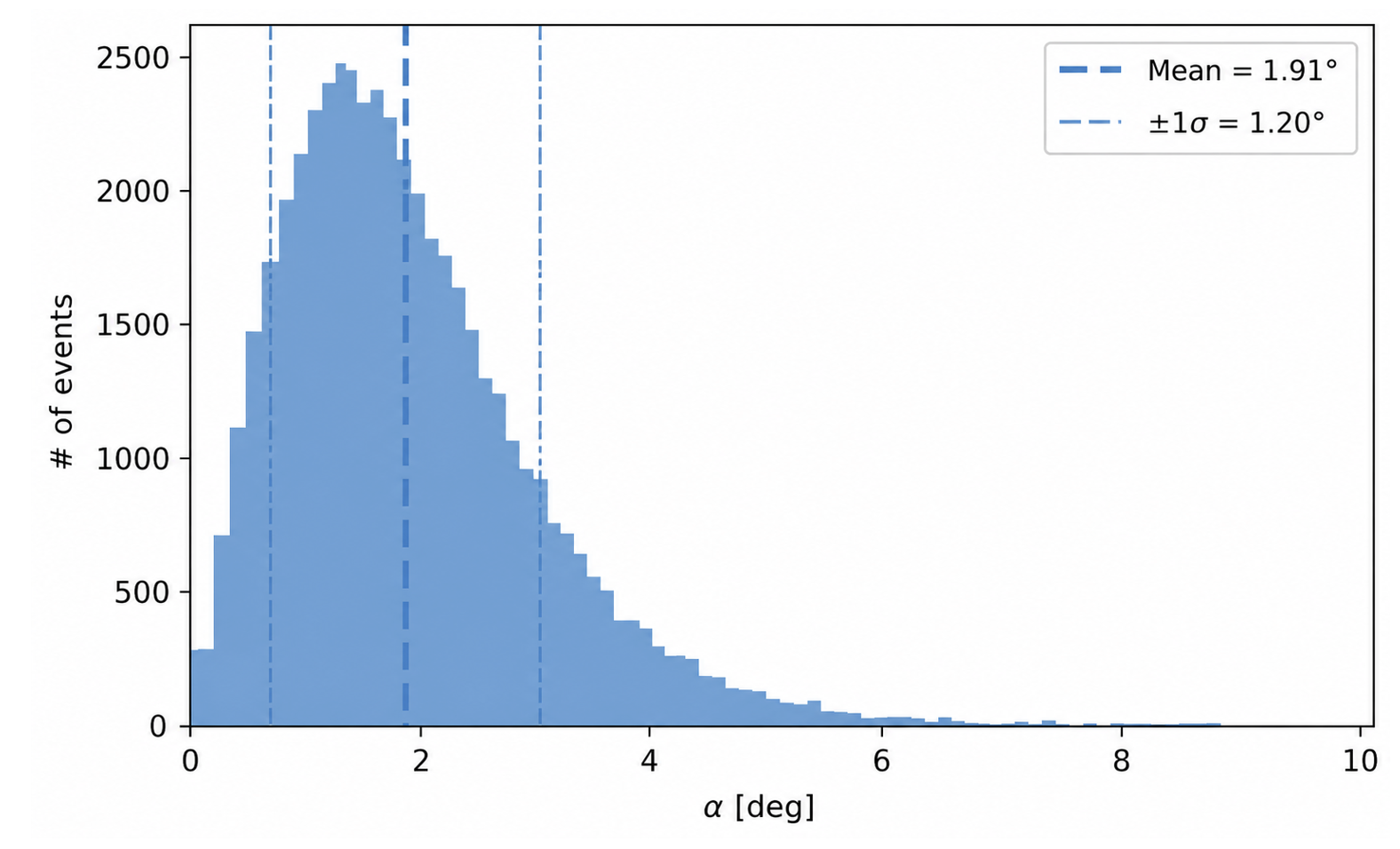}
        \caption{Triangular mesh reconstruction}
        \label{fig:hist_tri_simulations}
    \end{subfigure}
    \caption{Distribution of the angular separation, $\alpha$, obtained for 50,000 simulated GRBs reconstructed using the rectangular (a) and triangular (b) calibration meshes. The dashed lines indicate the mean angular separation and the $\pm1\sigma$ interval for each calibration geometry.}
    \label{fig:hist_simulations}
\end{figure}

The overall localization performance for the two calibration geometries is summarized in Figure \ref{fig:hist_simulations}, which shows the distribution of the angular separation, $\alpha$, from all 50,000 simulated GRBs. The triangular calibration mesh yields $\langle\alpha\rangle=1.91^\circ$ with $\sigma=1.20^\circ$. In contrast, the rectangular calibration mesh yields a mean angular separation of $\langle\alpha\rangle=2.73^\circ$ with $\sigma=2.18^\circ$. Both calibration geometries achieve stable localization performance under realistic space-observation conditions. However, the triangular calibration mesh, on average, yields smaller angular separations than the rectangular mesh.

Table~\ref{tab:localization_summary} summarizes the localization performance obtained for all calibration configurations investigated in this work, including laboratory measurements and simulations. The table lists the average angular separation, $\langle\alpha\rangle$, together with the corresponding standard deviation. The simulations differ from the laboratory measurements in three aspects. First, they include background; second, they implement a broadband GRB spectrum unlike the laboratory spectrum line at 59.6\,keV; third, the simulated flux is approximately 3--4 times smaller than that of the laboratory short 0.5\,s exposures. This explains the better laboratory localization results listed in Table~\ref{tab:localization_summary}. 

\begin{table}[h!]
\centering
\begin{tabular}{lllcc}
\hline
\textbf{Case Study} &
\textbf{Measurement Test} &
\textbf{Calibration Mesh} &
\textbf{$\langle\alpha\rangle$ [$^\circ$]} &
\textbf{$\sigma$ [$^\circ$]} \\
\hline
Laboratory & On-grid \cite{saleh2025} & Rectangular & 0.90 & 0.19 \\
Laboratory & Random off-grid & Rectangular & 1.19 & 0.28 \\
Laboratory & Random off-grid& Triangular & 0.86 & 0.14 \\
\hline
Simulation & Random off-grid & Rectangular & 2.73 & 2.18 \\
Simulation & Random off-grid & Triangular & 1.91 & 1.20 \\
\hline
\end{tabular}
\caption{\label{tab:localization_summary}
Summary of the localization performance obtained from the laboratory measurements and Monte Carlo simulations using the rectangular and triangular calibration meshes. The table lists the average angular separation, $\langle\alpha\rangle$, together with the corresponding standard deviation.
Note: The higher $\langle\alpha\rangle$ in the simulations is due to the inclusion of sky background and weaker flux.}
\end{table}

Overall, the triangular calibration mesh consistently yields smaller average angular separations and lower standard deviations than the rectangular mesh in both laboratory measurements and Monte Carlo simulations. These results demonstrate that the triangular calibration geometry provides a more accurate and consistent interpolation framework for GRB localization with GALI.

\subsection{Simulation Results of the Effective Area}
\label{subsec:effective_area}

\begin{figure}[h!]
    \centering
    \includegraphics[width=0.7\linewidth]{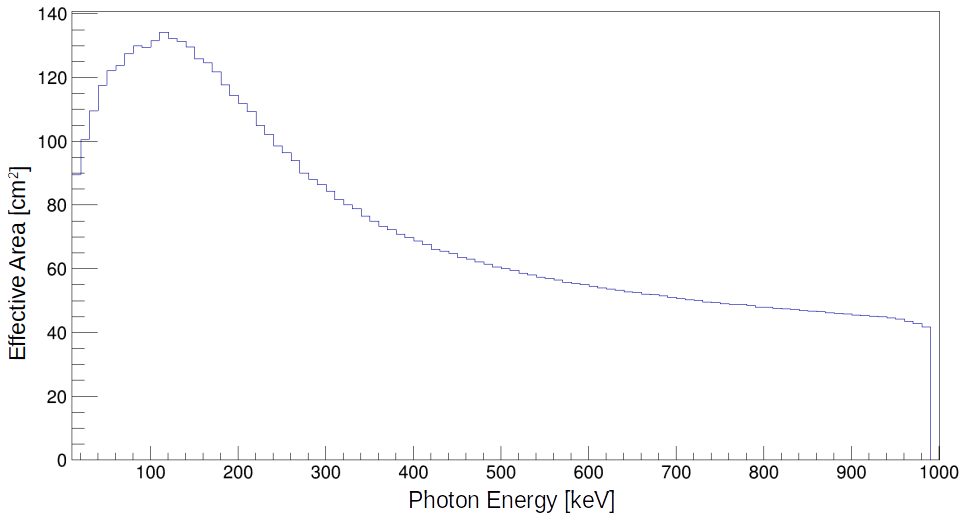}
    \caption{Simulated hemisphere-averaged effective area of the current GALI detector model as a function of photon energy. The simulations are performed for 70,232 GRBs uniformly distributed over the upper hemisphere. The effective area reaches a maximum of approximately 130\,cm$^2$ near 100--150\,keV and gradually decreases toward higher energies while maintaining sensitivity up to $\sim$ 1\,MeV.}
    \label{fig:effective_area_sim}
\end{figure}

\paragraph{} While the laboratory measurements presented in Figure \ref{fig:skymap_tri} provide the effective area of the current GALI detector model at 59.6\,keV, Monte Carlo simulations allow the detector response to be evaluated over the 10--1000\,keV energy range. To characterize the energy dependence of the detector sensitivity, 70,232 GRBs are simulated in vacuum without background radiation. The simulated bursts are uniformly distributed over the upper hemisphere, with $\sim$70 bursts generated at each photon-energy bin. For each simulation, the number of photons registered by the detector is recorded and used to determine the effective area. The effective area is then averaged over all source directions across the hemisphere and shown in Figure \ref{fig:effective_area_sim}.

The effective area reaches a maximum value of approximately 130\,cm$^2$ in the 100--150\,keV energy range, corresponding to the region of highest detection efficiency. This maximum lies within the 50--300\,keV energy band commonly used for GRB triggering, indicating that the current GALI detector design is well matched to the energy range where GRB detection is expected to be most efficient. At higher photon energies, the effective area gradually decreases because of the reduced probability of $\gamma$-ray interactions within the detector volume. Nevertheless, GALI maintains significant sensitivity across a broad energy range extending to $\sim$ 1\,MeV, supporting both efficient GRB detection and localization over a wide energy range. The laboratory effective area at 59.6\,keV is approximately 30\% smaller than the simulation value. The origin of this discrepancy is not yet fully understood. Further investigation will be carried out to determine whether this difference arises from detector efficiency, calibration effects, or other experimental and modeling uncertainties.

\section{Future Scalability and Mission Applications}
\label{sec:future}

\begin{figure}[h!]
\centering
\includegraphics[width=0.3\linewidth]{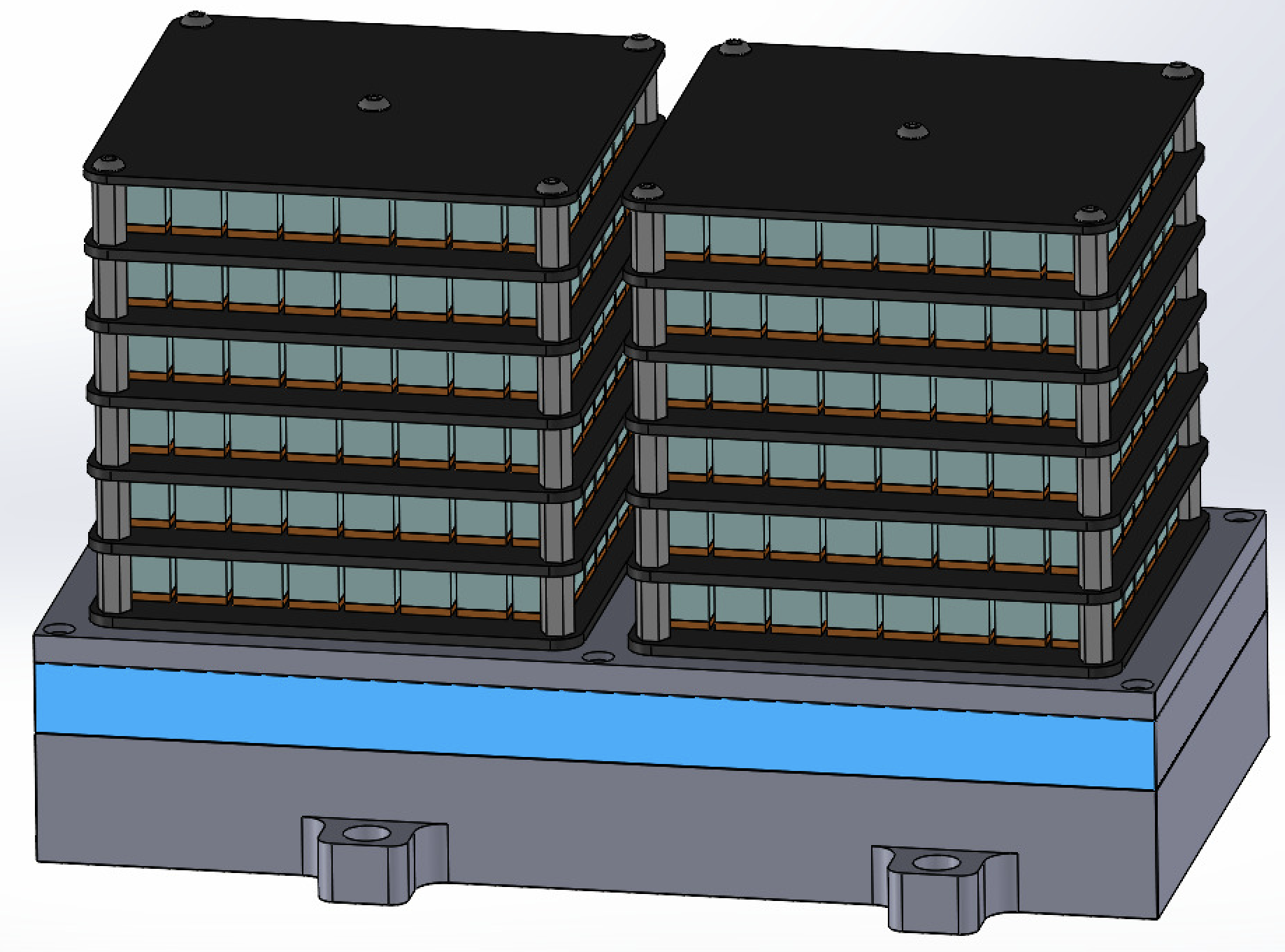}
\caption{Conceptual modular GALI architecture composed of multiple detector units. The modular design enables adaptation of the detector volume and effective area for future CubeSat, small-satellite, and ISS-based GRB localization missions while preserving the localization principle.}
\label{fig:new_GALI}
\end{figure}

\paragraph{} The modular architecture of GALI provides a flexible framework for developing detector configurations tailored to different mission requirements while preserving the localization principle. Compact implementations are well suited for CubeSat missions, where instrument volume, mass, and power consumption are tightly constrained. In contrast, multiple detector units can be combined to increase the effective area, improve photon statistics, and enhance localization performance. Figure~\ref{fig:new_GALI} illustrates a conceptual modular GALI architecture, including a 2U configuration composed of two detector units.

To evaluate the scalability of the GALI concept, Monte Carlo simulations are performed for both compact 1U and 2U CubeSat implementations. The detector models consist of randomly distributed CsI scintillators within an $8\times7\times6$ array with 126 scintillators, while the 2U configuration shown in Figure~\ref{fig:new_GALI} combines two identical detector units (252 scintillators). For each detector configuration, 12,000 GRBs with a duration of 5\,s and a photon flux of $10~\gamma~\mathrm{cm}^{-2}\,\mathrm{s}^{-1}$ are simulated. The 1U configuration achieves $\langle\alpha\rangle=2.52^\circ$ with $\sigma=1.97^\circ$, whereas the 2U configuration achieves $\langle\alpha\rangle=1.49^\circ$ with $\sigma=1.20^\circ$. These results demonstrate that GALI maintains degree-scale localization within the volume constraints of a single CubeSat unit and that combining two detector units significantly improves the localization performance while preserving the modular detector concept.

\section{Conclusions}
\label{sec:conclusions}

\paragraph{} This work presents and experimentally validates a triangular calibration mesh based on an icosphere tessellation for GALI localization. Compared with the conventional rectangular calibration mesh, the triangular geometry provides a more uniform sampling of the hemisphere while maintaining a similar calibration density, resulting in improved interpolation and localization performance.

Random off-grid laboratory measurements demonstrate that both calibration geometries achieve degree-scale localization. However, the triangular calibration mesh consistently yields better localization performance, achieving an average angular separation of $\langle\alpha_{\mathrm{avg}}\rangle = 0.86^\circ \pm 0.14^\circ$, compared with $\langle\alpha_{\mathrm{avg}}\rangle = 1.19^\circ \pm 0.28^\circ$ for the rectangular calibration mesh.

Monte Carlo simulations of 50,000 randomly distributed GRBs under Low Earth Orbit background conditions further validate the localization methodology. The simulations reproduce the trend observed in the laboratory measurements, with the triangular calibration mesh achieving an average angular separation of $\langle\alpha\rangle = 1.91^\circ \pm 1.20^\circ$, compared with $\langle\alpha\rangle = 2.73^\circ \pm 2.18^\circ$ for the rectangular calibration mesh. These results demonstrate robust degree-scale localization across the full hemisphere and confirm the advantage of the triangular calibration geometry under realistic observational conditions.

Effective area simulations show that the current GALI detector model remains sensitive over the 10\,keV to approximately 1\,MeV energy range, reaching a maximum effective area of approximately $130\,\mathrm{cm}^2$ near 100--150\,keV. Since this maximum falls within the 50--300\,keV energy band commonly used for GRB triggering, the detector is well matched to the energy range where GRB detection is expected to be most efficient.

Finally, Monte Carlo simulations of compact 1U and 2U GALI CubeSat configurations demonstrate the scalability of the modular detector concept. The 1U configuration achieves an average angular separation of $\langle\alpha\rangle = 2.52^\circ \pm 1.97^\circ$, while the 2U configuration improves the localization performance to $\langle\alpha\rangle = 1.49^\circ \pm 1.20^\circ$. These results indicate that GALI can maintain degree-scale localization within CubeSat volume constraints and that combining multiple detector units provides a straightforward path toward increased effective area and enhanced localization performance for future missions.

\acknowledgments
A grant from the Pazi Foundation and the Technion ADRI Research Institute supported this research. J.S.N. was supported by a fellowship from the Ministry of Innovation, Science and Technology, and by a scholarship awarded by the Alumni Unit of the Darca School Network.

% References
\bibliography{report} % bibliography data in report.bib

@ARTICLE{Meszaros2006,
       author = {{M{\'e}sz{\'a}ros}, P.},
        title = "{Gamma-ray bursts}",
      journal = {Reports on Progress in Physics},
         year = 2006,
        month = aug,
       volume = {69},
       number = {8},
        pages = {2259-2321},
          doi = {10.1088/0034-4885/69/8/R01},
archivePrefix = {arXiv},
       eprint = {astro-ph/0605208},
 primaryClass = {astro-ph},
       adsurl = {https://ui.adsabs.harvard.edu/abs/2006RPPh...69.2259M}
}

@article{Kumar2015,
       author = {{Kumar}, Pawan and {Zhang}, Bing},
        title = "{The physics of gamma-ray bursts \& relativistic jets}",
      journal = {physrep},
         year = 2015,
        month = feb,
       volume = {561},
        pages = {1-109},
          doi = {10.1016/j.physrep.2014.09.008},
archivePrefix = {arXiv},
       eprint = {1410.0679},
 primaryClass = {astro-ph.HE},
       adsurl = {https://ui.adsabs.harvard.edu/abs/2015PhR...561....1K}
}

@ARTICLE{Barthelmy2005,
       author = {{Barthelmy}, Scott D. and {Barbier}, Louis M. and {Cummings}, Jay R. and {Fenimore}, Ed E. and {Gehrels}, Neil and {Hullinger}, Derek and {Krimm}, Hans A. and {Markwardt}, Craig B. and {Palmer}, David M. and {Parsons}, Ann and {Sato}, Goro and {Suzuki}, Masaya and {Takahashi}, Tadayuki and {Tashiro}, Makota and {Tueller}, Jack},
        title = "{The Burst Alert Telescope (BAT) on the SWIFT Midex Mission}",
      journal = {ssr},
         year = 2005,
        month = oct,
       volume = {120},
       number = {3-4},
        pages = {143-164},
          doi = {10.1007/s11214-005-5096-3},
archivePrefix = {arXiv},
       eprint = {astro-ph/0507410},
 primaryClass = {astro-ph},
       adsurl = {https://ui.adsabs.harvard.edu/abs/2005SSRv..120..143B}
}

@ARTICLE{Fruchter2006,
       author = {{Fruchter}, A.~S. and {Levan}, A.~J. and {Strolger}, L. and {Vreeswijk}, P.~M. and {Thorsett}, S.~E. and {Bersier}, D. and {Burud}, I. and {Castro Cer{\'o}n}, J.~M. and {Castro-Tirado}, A.~J. and {Conselice}, C. and {Dahlen}, T. and {Ferguson}, H.~C. and {Fynbo}, J.~P.~U. and {Garnavich}, P.~M. and {Gibbons}, R.~A. and {Gorosabel}, J. and {Gull}, T.~R. and {Hjorth}, J. and {Holland}, S.~T. and {Kouveliotou}, C. and {Levay}, Z. and {Livio}, M. and {Metzger}, M.~R. and {Nugent}, P.~E. and {Petro}, L. and {Pian}, E. and {Rhoads}, J.~E. and {Riess}, A.~G. and {Sahu}, K.~C. and {Smette}, A. and {Tanvir}, N.~R. and {Wijers}, R.~A.~M.~J. and {Woosley}, S.~E.},
        title = "{Long {\ensuremath{\gamma}}-ray bursts and core-collapse supernovae have different environments}",
      journal = {nat},
         year = 2006,
        month = may,
       volume = {441},
       number = {7092},
        pages = {463-468},
          doi = {10.1038/nature04787},
archivePrefix = {arXiv},
       eprint = {astro-ph/0603537},
 primaryClass = {astro-ph},
       adsurl = {https://ui.adsabs.harvard.edu/abs/2006Natur.441..463F}
}

@article{Abbott2017a,
       author = {{Abbott}, B.~P. and {Abbott}, R. and {Abbott}, T.~D. and {Acernese}, F. and {Ackley}, K. and {Adams}, C. and {Adams}, T. and {Addesso}, P. and {Adhikari}, R.~X. and {Adya}, V.~B. and {Affeldt}, C. and {Afrough}, M. and {Agarwal}, B. and {Agathos}, M. and {Agatsuma}, K. and {Aggarwal}, N. and {Aguiar}, O.~D. and {Aiello}, L. and {Ain}, A. and {Ajith}, P. and {Allen}, B. and {Allen}, G. and {Allocca}, A. and {Aloy}, M.~A. and {Altin}, P.~A. and {Amato}, A. and {Ananyeva}, A. and {Anderson}, S.~B. and {Anderson}, W.~G. and {Angelova}, S.~V. and {Antier}, S. and {Appert}, S. and {Arai}, K. and {Araya}, M.~C. and {Areeda}, J.~S. and {Arnaud}, N. and {Arun}, K.~G. and {Ascenzi}, S. and {Ashton}, G. and {Ast}, M. and {Aston}, S.~M. and {Astone}, P. and {Atallah}, D.~V. and {Aufmuth}, P. and {Aulbert}, C. and {AultONeal}, K. and {Austin}, C. and {Avila-Alvarez}, A. and {Babak}, S. and {Bacon}, P. and {Bader}, M.~K.~M. and {Bae}, S. and {Baker}, P.~T. and {Baldaccini}, F. and {Ballardin}, G. and {Ballmer}, S.~W. and {Banagiri}, S. and {Barayoga}, J.~C. and {Barclay}, S.~E. and {Barish}, B.~C. and {Barker}, D. and {Barkett}, K. and {Barone}, F. and {Barr}, B. and {Barsotti}, L. and {Barsuglia}, M. and {Barta}, D. and {Bartlett}, J. and {Bartos}, I. and {Bassiri}, R. and {Basti}, A. and {Batch}, J.~C. and {Bawaj}, M. and {Bayley}, J.~C. and {Bazzan}, M. and {B{\'e}csy}, B. and {Beer}, C. and {Bejger}, M. and {Belahcene}, I. and {Bell}, A.~S. and {Berger}, B.~K. and {Bergmann}, G. and {Bero}, J.~J. and {Berry}, C.~P.~L. and {Bersanetti}, D. and {Bertolini}, A. and {Betzwieser}, J. and {Bhagwat}, S. and {Bhandare}, R. and {Bilenko}, I.~A. and {Billingsley}, G. and {Billman}, C.~R. and {Birch}, J. and {Birney}, R. and {Birnholtz}, O. and {Biscans}, S. and {Biscoveanu}, S. and {Bisht}, A. and {Bitossi}, M. and {Biwer}, C. and {Bizouard}, M.~A. and {Blackburn}, J.~K. and {Blackman}, J. and {Blair}, C.~D. and {Blair}, D.~G. and {Blair}, R.~M. and {Bloemen}, S. and {Bock}, O. and {Bode}, N. and {Boer}, M. and {Bogaert}, G. and {Bohe}, A. and {Bondu}, F. and {Bonilla}, E. and {Bonnand}, R. and {Boom}, B.~A. and {Bork}, R. and {Boschi}, V. and {Bose}, S. and {Bossie}, K. and {Bouffanais}, Y. and {Bozzi}, A. and {Bradaschia}, C. and {Brady}, P.~R. and {Branchesi}, M. and {Brau}, J.~E. and {Briant}, T. and {Brillet}, A. and {Brinkmann}, M. and {Brisson}, V. and {Brockill}, P. and {Broida}, J.~E. and {Brooks}, A.~F. and {Brown}, D.~A. and {Brown}, D.~D. and {Brunett}, S. and {Buchanan}, C.~C. and {Buikema}, A. and {Bulik}, T. and {Bulten}, H.~J. and {Buonanno}, A. and {Buskulic}, D. and {Buy}, C. and {Byer}, R.~L. and {Cabero}, M. and {Cadonati}, L. and {Cagnoli}, G. and {Cahillane}, C. and {Calder{\'o}n Bustillo}, J. and {Callister}, T.~A. and {Calloni}, E. and {Camp}, J.~B. and {Canepa}, M. and {Canizares}, P. and {Cannon}, K.~C. and {Cao}, H. and {Cao}, J. and {Capano}, C.~D. and {Capocasa}, E. and {Carbognani}, F. and {Caride}, S. and {Carney}, M.~F. and {Casanueva Diaz}, J. and {Casentini}, C. and {Caudill}, S. and {Cavagli{\`a}}, M. and {Cavalier}, F. and {Cavalieri}, R. and {Cella}, G. and {Cepeda}, C.~B. and {Cerd{\'a}-Dur{\'a}n}, P. and {Cerretani}, G. and {Cesarini}, E. and {Chamberlin}, S.~J. and {Chan}, M. and {Chao}, S. and {Charlton}, P. and {Chase}, E. and {Chassande-Mottin}, E. and {Chatterjee}, D. and {Chatziioannou}, K. and {Cheeseboro}, B.~D. and {Chen}, H.~Y. and {Chen}, X. and {Chen}, Y. and {Cheng}, H. -P. and {Chia}, H. and {Chincarini}, A. and {Chiummo}, A. and {Chmiel}, T. and {Cho}, H.~S. and {Cho}, M. and {Chow}, J.~H. and {Christensen}, N. and {Chu}, Q. and {Chua}, A.~J.~K. and {Chua}, S. and {Chung}, A.~K.~W. and {Chung}, S. and {Ciani}, G.},
        title = "{Gravitational Waves and Gamma-Rays from a Binary Neutron Star Merger: GW170817 and GRB 170817A}",
      journal = {apjl},
         year = 2017,
        month = oct,
       volume = {848},
       number = {2},
          eid = {L13},
        pages = {L13},
          doi = {10.3847/2041-8213/aa920c},
archivePrefix = {arXiv},
       eprint = {1710.05834},
 primaryClass = {astro-ph.HE},
       adsurl = {https://ui.adsabs.harvard.edu/abs/2017ApJ...848L..13A}
}

@ARTICLE{Meegan2009,
       author = {{Meegan}, Charles and {Lichti}, Giselher and {Bhat}, P.~N. and {Bissaldi}, Elisabetta and {Briggs}, Michael S. and {Connaughton}, Valerie and {Diehl}, Roland and {Fishman}, Gerald and {Greiner}, Jochen and {Hoover}, Andrew S. and {van der Horst}, Alexander J. and {von Kienlin}, Andreas and {Kippen}, R. Marc and {Kouveliotou}, Chryssa and {McBreen}, Sheila and {Paciesas}, W.~S. and {Preece}, Robert and {Steinle}, Helmut and {Wallace}, Mark S. and {Wilson}, Robert B. and {Wilson-Hodge}, Colleen},
        title = "{The Fermi Gamma-ray Burst Monitor}",
      journal = {apj},
         year = 2009,
        month = sep,
       volume = {702},
       number = {1},
        pages = {791-804},
          doi = {10.1088/0004-637X/702/1/791},
archivePrefix = {arXiv},
       eprint = {0908.0450},
 primaryClass = {astro-ph.IM},
       adsurl = {https://ui.adsabs.harvard.edu/abs/2009ApJ...702..791M}
}

@article{ubertini2003,
  title={IBIS: The Imager on-board INTEGRAL},
  author={Ubertini, P and Lebrun, F and Di Cocco, Gea and Bazzano, A and Bird, AJ and Broenstad, K ea and Goldwurm, A and La Rosa, G and Labanti, C and Laurent, P and others},
  journal={Astronomy \& Astrophysics},
  volume={411},
  number={1},
  pages={L131--L139},
  year={2003},
  publisher={EDP Sciences}
}

@ARTICLE{Yuan2015,
       author = {{Yuan}, Weimin and {Zhang}, C. and {Feng}, H. and {Zhang}, S.~N. and {Ling}, Z.~X. and {Zhao}, D. and {Deng}, J. and {Qiu}, Y. and {Osborne}, J.~P. and {O'Brien}, P. and {Willingale}, R. and {Lapington}, J. and {Fraser}, G.~W. and {the Einstein Probe team}},
        title = "{Einstein Probe - a small mission to monitor and explore the dynamic X-ray Universe}",
      journal = {arXiv e-prints},
         year = 2015,
        month = jun,
          eid = {arXiv:1506.07735},
        pages = {arXiv:1506.07735},
          doi = {10.48550/arXiv.1506.07735},
archivePrefix = {arXiv},
       eprint = {1506.07735},
 primaryClass = {astro-ph.HE},
       adsurl = {https://ui.adsabs.harvard.edu/abs/2015arXiv150607735Y}
}

@ARTICLE{Amati2018,
       author = {{Amati}, L. and {O'Brien}, P. and {G{\"o}tz}, D. and {Bozzo}, E. and {Tenzer}, C. and {Frontera}, F. and {Ghirlanda}, G. and {Labanti}, C. and {Osborne}, J.~P. and {Stratta}, G. and {Tanvir}, N. and {Willingale}, R. and {Attina}, P. and {Campana}, R. and {Castro-Tirado}, A.~J. and {Contini}, C. and {Fuschino}, F. and {Gomboc}, A. and {Hudec}, R. and {Orleanski}, P. and {Renotte}, E. and {Rodic}, T. and {Bagoly}, Z. and {Blain}, A. and {Callanan}, P. and {Covino}, S. and {Ferrara}, A. and {Le Floch}, E. and {Marisaldi}, M. and {Mereghetti}, S. and {Rosati}, P. and {Vacchi}, A. and {D'Avanzo}, P. and {Giommi}, P. and {Piranomonte}, S. and {Piro}, L. and {Reglero}, V. and {Rossi}, A. and {Santangelo}, A. and {Salvaterra}, R. and {Tagliaferri}, G. and {Vergani}, S. and {Vinciguerra}, S. and {Briggs}, M. and {Campolongo}, E. and {Ciolfi}, R. and {Connaughton}, V. and {Cordier}, B. and {Morelli}, B. and {Orlandini}, M. and {Adami}, C. and {Argan}, A. and {Atteia}, J.-L. and {Auricchio}, N. and {Balazs}, L. and {Baldazzi}, G. and {Basa}, S. and {Basak}, R. and {Bellutti}, P. and {Bernardini}, M.~G. and {Bertuccio}, G. and {Braga}, J. and {Branchesi}, M. and {Brandt}, S. and {Brocato}, E. and {Budtz-Jorgensen}, C. and {Bulgarelli}, A. and {Burderi}, L. and {Camp}, J. and {Capozziello}, S. and {Caruana}, J. and {Casella}, P. and {Cenko}, B. and {Chardonnet}, P. and {Ciardi}, B. and {Colafrancesco}, S. and {Dainotti}, M.~G. and {D'Elia}, V. and {De Martino}, D. and {De Pasquale}, M. and {Del Monte}, E. and {Della Valle}, M. and {Drago}, A. and {Evangelista}, Y. and {Feroci}, M. and {Finelli}, F. and {Fiorini}, M. and {Fynbo}, J. and {Gal-Yam}, A. and {Gendre}, B. and {Ghisellini}, G. and {Grado}, A. and {Guidorzi}, C. and {Hafizi}, M. and {Hanlon}, L. and {Hjorth}, J. and {Izzo}, L. and {Kiss}, L. and {Kumar}, P. and {Kuvvetli}, I. and {Lavagna}, M. and {Li}, T. and {Longo}, F. and {Lyutikov}, M. and {Maio}, U. and {Maiorano}, E. and {Malcovati}, P. and {Malesani}, D. and {Margutti}, R. and {Martin-Carrillo}, A. and {Masetti}, N. and {McBreen}, S. and {Mignani}, R. and {Morgante}, G. and {Mundell}, C. and {Nargaard-Nielsen}, H.~U. and {Nicastro}, L. and {Palazzi}, E. and {Paltani}, S. and {Panessa}, F. and {Pareschi}, G. and {Pe'er}, A. and {Penacchioni}, A.~V. and {Pian}, E. and {Piedipalumbo}, E. and {Piran}, T. and {Rauw}, G. and {Razzano}, M. and {Read}, A. and {Rezzolla}, L. and {Romano}, P. and {Ruffini}, R. and {Savaglio}, S. and {Sguera}, V. and {Schady}, P. and {Skidmore}, W. and {Song}, L. and {Stanway}, E. and {Starling}, R. and {Topinka}, M. and {Troja}, E. and {van Putten}, M. and {Vanzella}, E. and {Vercellone}, S. and {Wilson-Hodge}, C. and {Yonetoku}, D. and {Zampa}, G. and {Zampa}, N. and {Zhang}, B. and {Zhang}, B.~B. and {Zhang}, S. and {Zhang}, S.-N. and {Antonelli}, A. and {Bianco}, F. and {Boci}, S. and {Boer}, M. and {Botticella}, M.~T. and {Boulade}, O. and {Butler}, C. and {Campana}, S. and {Capitanio}, F. and {Celotti}, A. and {Chen}, Y. and {Colpi}, M. and {Comastri}, A. and {Cuby}, J.-G. and {Dadina}, M. and {De Luca}, A. and {Dong}, Y.-W. and {Ettori}, S. and {Gandhi}, P. and {Geza}, E. and {Greiner}, J. and {Guiriec}, S. and {Harms}, J. and {Hernanz}, M. and {Hornstrup}, A. and {Hutchinson}, I. and {Israel}, G. and {Jonker}, P. and {Kaneko}, Y. and {Kawai}, N. and {Wiersema}, K. and {Korpela}, S. and {Lebrun}, V. and {Lu}, F. and {MacFadyen}, A. and {Malaguti}, G. and {Maraschi}, L. and {Melandri}, A. and {Modjaz}, M. and {Morris}, D. and {Omodei}, N. and {Paizis}, A. and {P{\'a}ta}, P. and {Petrosian}, V. and {Rachevski}, A. and {Rhoads}, J. and {Ryde}, F. and {Sabau-Graziati}, L.},
        title = "{The THESEUS space mission concept: science case, design and expected performances}",
      journal = {Advances in Space Research},
         year = 2018,
        month = jul,
       volume = {62},
       number = {1},
        pages = {191-244},
          doi = {10.1016/j.asr.2018.03.010},
archivePrefix = {arXiv},
       eprint = {1710.04638},
 primaryClass = {astro-ph.IM},
       adsurl = {https://ui.adsabs.harvard.edu/abs/2018AdSpR..62..191A}
}

@ARTICLE{Wei2016,
       author = {{Wei}, J. and {Cordier}, B. and {Antier}, S. and {Antilogus}, P. and {Atteia}, J.-L. and {Bajat}, A. and {Basa}, S. and {Beckmann}, V. and {Bernardini}, M.~G. and {Boissier}, S. and {Bouchet}, L. and {Burwitz}, V. and {Claret}, A. and {Dai}, Z.-G. and {Daigne}, F. and {Deng}, J. and {Dornic}, D. and {Feng}, H. and {Foglizzo}, T. and {Gao}, H. and {Gehrels}, N. and {Godet}, O. and {Goldwurm}, A. and {Gonzalez}, F. and {Gosset}, L. and {G{\"o}tz}, D. and {Gouiffes}, C. and {Grise}, F. and {Gros}, A. and {Guilet}, J. and {Han}, X. and {Huang}, M. and {Huang}, Y.-F. and {Jouret}, M. and {Klotz}, A. and {La Marle}, O. and {Lachaud}, C. and {Le Floch}, E. and {Lee}, W. and {Leroy}, N. and {Li}, L.-X. and {Li}, S.~C. and {Li}, Z. and {Liang}, E.-W. and {Lyu}, H. and {Mercier}, K. and {Migliori}, G. and {Mochkovitch}, R. and {O'Brien}, P. and {Osborne}, J. and {Paul}, J. and {Perinati}, E. and {Petitjean}, P. and {Piron}, F. and {Qiu}, Y. and {Rau}, A. and {Rodriguez}, J. and {Schanne}, S. and {Tanvir}, N. and {Vangioni}, E. and {Vergani}, S. and {Wang}, F.-Y. and {Wang}, J. and {Wang}, X.-G. and {Wang}, X.-Y. and {Watson}, A. and {Webb}, N. and {Wei}, J.~J. and {Willingale}, R. and {Wu}, C. and {Wu}, X.-F. and {Xin}, L.-P. and {Xu}, D. and {Yu}, S. and {Yu}, W.-F. and {Yu}, Y.-W. and {Zhang}, B. and {Zhang}, S.-N. and {Zhang}, Y. and {Zhou}, X.~L.},
        title = "{The Deep and Transient Universe in the SVOM Era: New Challenges and Opportunities - Scientific prospects of the SVOM mission}",
      journal = {arXiv e-prints},
         year = 2016,
        month = oct,
          eid = {arXiv:1610.06892},
        pages = {arXiv:1610.06892},
          doi = {10.48550/arXiv.1610.06892},
archivePrefix = {arXiv},
       eprint = {1610.06892},
 primaryClass = {astro-ph.IM},
       adsurl = {https://ui.adsabs.harvard.edu/abs/2016arXiv161006892W}
}

@inproceedings{sanna2020,
  title={Timing techniques applied to distributed modular high-energy astronomy: the HERMES project},
  author={Sanna, Andrea and Burderi, Luciano and Di Salvo, Tiziana and Fiore, Fabrizio and Riggio, Alessandro and Gambino, Angelo and Lavagna, Mich{\`e}le and Bertacin, Roberto and Evangelista, Yuri and Campana, Riccardo and others},
  booktitle={Space Telescopes and Instrumentation 2020: Ultraviolet to Gamma Ray},
  volume={11444},
  pages={844--862},
  year={2020},
  organization={SPIE}
}

@ARTICLE{saleh2025,
       author = {{Saleh-Natur}, Julia and {Behar}, Ehud and {Reich}, Omer and {Tarem}, Shlomit and {Tarem}, Zvika and {Vdovin}, Alexander and {Feigenboim}, Amir and {Avioz}, Liron and {Klimov}, Aleksei and {Mitrani}, Sharon and {Margolin}, Solomon and {Kaidar}, Avner and {Agalarian}, Hovhannes and {Rahin}, Roi and {Osovizky}, Alon and {Vax}, Eran and {Ghelman}, Max},
        title = "{GALI {\textemdash} A GAmma-ray burst Localizing Instrument}",
      journal = {Nuclear Instruments and Methods in Physics Research A},
         year = 2026,
        month = apr,
       volume = {1084},
          eid = {171226},
        pages = {171226},
          doi = {10.1016/j.nima.2025.171226},
       adsurl = {https://ui.adsabs.harvard.edu/abs/2026NIMPA108471226S}
}

@article{Cash1979,
       author = {{Cash}, W.},
        title = "{Parameter estimation in astronomy through application of the likelihood ratio.}",
      journal = {apj},
         year = 1979,
        month = mar,
       volume = {228},
        pages = {939-947},
          doi = {10.1086/156922},
       adsurl = {https://ui.adsabs.harvard.edu/abs/1979ApJ...228..939C}
}

@ARTICLE{Lee2017,
       author = {{Lee}, Sanghyun and {Mortari}, Daniele},
        title = "{Quasi-equal area subdivision algorithm for uniform points on a sphere with application to any geographical data distribution}",
      journal = {Computers and Geosciences},
         year = 2017,
        month = jun,
       volume = {103},
        pages = {142-151},
          doi = {10.1016/j.cageo.2017.03.012},
       adsurl = {https://ui.adsabs.harvard.edu/abs/2017CG....103..142L}
}

@inproceedings{saleh2024,
  title={GALI-a GAmma-ray burst Localizing Instrument: results from full-size engineering model},
  author={Saleh-Natur, Julia and Behar, Ehud and Reich, Omer and Tarem, Shlomit and Tarem, Zvika and Vdovin, Alex and Feigenboim, Amir and Rahin, Roi and Kaidar, Avner and Agalarian, Hovannes and others},
  booktitle={Space Telescopes and Instrumentation 2024: Ultraviolet to Gamma Ray},
  volume={13093},
  pages={2171--2181},
  year={2024},
  organization={SPIE}
}

@software{Zoglauer.soft06018Z,
       author = {{Zoglauer}, Andreas},
        title = "{MEGAlib: Medium Energy Gamma-ray Astronomy library}",
 howpublished = {Astrophysics Source Code Library, record ascl:1906.018},
         year = 2019,
        month = jun,
          eid = {ascl:1906.018},
       adsurl = {https://ui.adsabs.harvard.edu/abs/2019ascl.soft06018Z}
}

@article{Agostinelli2003,
title = {Geant4—a simulation toolkit},
journal = {Nuclear Instruments and Methods in Physics Research Section A: Accelerators, Spectrometers, Detectors and Associated Equipment},
volume = {506},
number = {3},
pages = {250-303},
year = {2003},
issn = {0168-9002},
doi = {https://doi.org/10.1016/S0168-9002(03)01368-8},
url = {https://www.sciencedirect.com/science/article/pii/S0168900203013688},
author = {S. Agostinelli and J. Allison and K. Amako and J. Apostolakis and H. Araujo and P. Arce and M. Asai and D. Axen and S. Banerjee and G. Barrand and F. Behner and L. Bellagamba and J. Boudreau and L. Broglia and A. Brunengo and H. Burkhardt and S. Chauvie and J. Chuma and R. Chytracek and G. Cooperman and G. Cosmo and P. Degtyarenko and A. Dell'Acqua and G. Depaola and D. Dietrich and R. Enami and A. Feliciello and C. Ferguson and H. Fesefeldt and G. Folger and F. Foppiano and A. Forti and S. Garelli and S. Giani and R. Giannitrapani and D. Gibin and J.J. {Gómez Cadenas} and I. González and G. {Gracia Abril} and G. Greeniaus and W. Greiner and V. Grichine and A. Grossheim and S. Guatelli and P. Gumplinger and R. Hamatsu and K. Hashimoto and H. Hasui and A. Heikkinen and A. Howard and V. Ivanchenko and A. Johnson and F.W. Jones and J. Kallenbach and N. Kanaya and M. Kawabata and Y. Kawabata and M. Kawaguti and S. Kelner and P. Kent and A. Kimura and T. Kodama and R. Kokoulin and M. Kossov and H. Kurashige and E. Lamanna and T. Lampén and V. Lara and V. Lefebure and F. Lei and M. Liendl and W. Lockman and F. Longo and S. Magni and M. Maire and E. Medernach and K. Minamimoto and P. {Mora de Freitas} and Y. Morita and K. Murakami and M. Nagamatu and R. Nartallo and P. Nieminen and T. Nishimura and K. Ohtsubo and M. Okamura and S. O'Neale and Y. Oohata and K. Paech and J. Perl and A. Pfeiffer and M.G. Pia and F. Ranjard and A. Rybin and S. Sadilov and E. {Di Salvo} and G. Santin and T. Sasaki and N. Savvas and Y. Sawada and S. Scherer and S. Sei and V. Sirotenko and D. Smith and N. Starkov and H. Stoecker and J. Sulkimo and M. Takahata and S. Tanaka and E. Tcherniaev and E. {Safai Tehrani} and M. Tropeano and P. Truscott and H. Uno and L. Urban and P. Urban and M. Verderi and A. Walkden and W. Wander and H. Weber and J.P. Wellisch and T. Wenaus and D.C. Williams and D. Wright and T. Yamada and H. Yoshida and D. Zschiesche},

}

@article{Allison2006,
       author = {{Allison}, J. and {Amako}, K. and {Apostolakis}, J. and {Araujo}, H. and {Dubois}, P. Arce and {Asai}, M. and {Barrand}, G. and {Capra}, R. and {Chauvie}, S. and {Chytracek}, R. and {Cirrone}, G.~A.~P. and {Cooperman}, G. and {Cosmo}, G. and {Cuttone}, G. and {Daquino}, G.~G. and {Donszelmann}, M. and {Dressel}, M. and {Folger}, G. and {Foppiano}, F. and {Generowicz}, J. and {Grichine}, V. and {Guatelli}, S. and {Gumplinger}, P. and {Heikkinen}, A. and {Hrivnacova}, I. and {Howard}, A. and {Incerti}, S. and {Ivanchenko}, V. and {Johnson}, T. and {Jones}, F. and {Koi}, T. and {Kokoulin}, R. and {Kossov}, M. and {Kurashige}, H. and {Lara}, V. and {Larsson}, S. and {Lei}, F. and {Link}, O. and {Longo}, F. and {Maire}, M. and {Mantero}, A. and {Mascialino}, B. and {McLaren}, I. and {Lorenzo}, P. Mendez and {Minamimoto}, K. and {Murakami}, K. and {Nieminen}, P. and {Pandola}, L. and {Parlati}, S. and {Peralta}, L. and {Perl}, J. and {Pfeiffer}, A. and {Pia}, M.~G. and {Ribon}, A. and {Rodrigues}, P. and {Russo}, G. and {Sadilov}, S. and {Santin}, G. and {Sasaki}, T. and {Smith}, D. and {Starkov}, N. and {Tanaka}, S. and {Tcherniaev}, E. and {Tom{\'e}}, B. and {Trindade}, A. and {Truscott}, P. and {Urban}, L. and {Verderi}, M. and {Walkden}, A. and {Wellisch}, J.~P. and {Williams}, D.~C. and {Wright}, D. and {Yoshida}, H.},
        title = "{Geant4 Developments and Applications}",
      journal = {IEEE Transactions on Nuclear Science},
         year = 2006,
        month = jan,
       volume = {53},
       number = {1},
        pages = {270-278},
          doi = {10.1109/TNS.2006.869826},
       adsurl = {https://ui.adsabs.harvard.edu/abs/2006ITNS...53..270A}
}

@article{Allison2016,
title = {Recent developments in Geant4},
journal = {Nuclear Instruments and Methods in Physics Research Section A: Accelerators, Spectrometers, Detectors and Associated Equipment},
volume = {835},
pages = {186-225},
year = {2016},
issn = {0168-9002},
doi = {https://doi.org/10.1016/j.nima.2016.06.125},
url = {https://www.sciencedirect.com/science/article/pii/S0168900216306957},
author = {J. Allison and K. Amako and J. Apostolakis and P. Arce and M. Asai and T. Aso and E. Bagli and A. Bagulya and S. Banerjee and G. Barrand and B.R. Beck and A.G. Bogdanov and D. Brandt and J.M.C. Brown and H. Burkhardt and Ph. Canal and D. Cano-Ott and S. Chauvie and K. Cho and G.A.P. Cirrone and G. Cooperman and M.A. Cortés-Giraldo and G. Cosmo and G. Cuttone and G. Depaola and L. Desorgher and X. Dong and A. Dotti and V.D. Elvira and G. Folger and Z. Francis and A. Galoyan and L. Garnier and M. Gayer and K.L. Genser and V.M. Grichine and S. Guatelli and P. Guèye and P. Gumplinger and A.S. Howard and I. Hřivnáčová and S. Hwang and S. Incerti and A. Ivanchenko and V.N. Ivanchenko and F.W. Jones and S.Y. Jun and P. Kaitaniemi and N. Karakatsanis and M. Karamitros and M. Kelsey and A. Kimura and T. Koi and H. Kurashige and A. Lechner and S.B. Lee and F. Longo and M. Maire and D. Mancusi and A. Mantero and E. Mendoza and B. Morgan and K. Murakami and T. Nikitina and L. Pandola and P. Paprocki and J. Perl and I. Petrović and M.G. Pia and W. Pokorski and J.M. Quesada and M. Raine and M.A. Reis and A. Ribon and A. {Ristić Fira} and F. Romano and G. Russo and G. Santin and T. Sasaki and D. Sawkey and J.I. Shin and I.I. Strakovsky and A. Taborda and S. Tanaka and B. Tomé and T. Toshito and H.N. Tran and P.R. Truscott and L. Urban and V. Uzhinsky and J.M. Verbeke and M. Verderi and B.L. Wendt and H. Wenzel and D.H. Wright and D.M. Wright and T. Yamashita and J. Yarba and H. Yoshida},
}

@ARTICLE{band1993,
       author = {{Band}, D. and {Matteson}, J. and {Ford}, L. and {Schaefer}, B. and {Palmer}, D. and {Teegarden}, B. and {Cline}, T. and {Briggs}, M. and {Paciesas}, W. and {Pendleton}, G. and {Fishman}, G. and {Kouveliotou}, C. and {Meegan}, C. and {Wilson}, R. and {Lestrade}, P.},
        title = "{BATSE Observations of Gamma-Ray Burst Spectra. I. Spectral Diversity}",
      journal = {apj},
         year = 1993,
        month = aug,
       volume = {413},
        pages = {281},
          doi = {10.1086/172995},
       adsurl = {https://ui.adsabs.harvard.edu/abs/1993ApJ...413..281B}
}
\bibliographystyle{spiebib} % makes bibtex use spiebib.bst

\end{document}